\documentclass[epj,nopacs]{svjour}

\usepackage{epsfig}
\usepackage{graphicx}
\usepackage{rotating}
\usepackage{color}

\newcommand{\drbar}{{\overline{\rm DR}}}
\newcommand{\msbar}{{\overline{\rm MS}}}
\newcommand{\GeV}{{\rm GeV}}
\newcommand{\TeV}{{\rm TeV}}

\newcommand{\stau}{\tilde{\tau}}
\newcommand{\snt}{{\tilde{\nu}_\tau}}
\newcommand{\ur}{\tilde{u}_R}
\newcommand{\ul}{\tilde{u}_L}
\newcommand{\dr}{\tilde{d}_R}
\newcommand{\dl}{\tilde{d}_L}
\newcommand{\st}{\tilde{t}}
\newcommand{\sbot}{\tilde{b}}
\newcommand{\sg}{\tilde{g}}
\newcommand{\nt}{\tilde{\chi}^0}
\newcommand{\cp}{\tilde{\chi}^+}
\newcommand{\cm}{\tilde{\chi}^-}
\newcommand{\cx}{\tilde{\chi}}
\newcommand{\ser}{\tilde{e}_R}
\newcommand{\sel}{\tilde{e}_L}
\newcommand{\sne}{\tilde{\nu}_e}

\makeatletter
\def\fmslash{\@ifnextchar[{\fmsl@sh}{\fmsl@sh[0mu]}}
\def\fmsl@sh[#1]#2{%
  \mathchoice
    {\@fmsl@sh\displaystyle{#1}{#2}}%
    {\@fmsl@sh\textstyle{#1}{#2}}%
    {\@fmsl@sh\scriptstyle{#1}{#2}}%
    {\@fmsl@sh\scriptscriptstyle{#1}{#2}}}
\def\@fmsl@sh#1#2#3{\m@th\ooalign{$\hfil#1\mkern#2/\hfil$\crcr$#1#3$}}
\makeatother

\def\makeheadbox{{%
\vspace*{-1\baselineskip}
\begin{flushright}
CERN--PH--TH/2005--232\\
DESY 05--242\\
FERMILAB--PUB--05--524--T\\
KEK--TH--1054\\
SLAC--PUB--11579
\end{flushright}
}}

\begin{document}
\hugehead

\title{Supersymmetry Parameter Analysis: SPA Convention \\
       and Project}

\titlerunning{Supersymmetry Parameter Analysis}
\authorrunning{J.A.~Aguilar-Saavedra {\it et al.} } 


\author{
J.A.~Aguilar-Saavedra$^{1}$, 
A.~Ali$^{ 2}$,               
B.C.~Allanach$^{ 3}$,        
R.~Arnowitt$^{ 4}$,          
H.A.~Baer$^{ 5}$,            
J.A.~Bagger$^{ 6}$,          
C.~Balazs$^{ 7\,a}$,         
V.~Barger$^{ 8}$,            
M.~Barnett$^{ 9}$,           
A.~Bartl$^{10}$
% ...
} 

\institute{ \it
  $ ^{~1}$ Departamento de Fisica and CFTP, 
           Instituto Superior Tecnico, Lisbon, Portugal \\ 
  $ ^{~2}$ Deutsches Elektronen-Synchrotron DESY, Hamburg, Germany \\
  $ ^{~3}$ DAMTP, University of Cambridge, Cambridge, UK \\
  $ ^{~4}$ Department of Physics, Texas A\&M University,
           College Station, TX, USA \\
  $ ^{~5}$ Department of Physics, Florida State University, Tallahassee, FL, USA \\
  $ ^{~6}$ Department of Physics and Astronomy, Johns Hopkins University, 
           Baltimore, MD, USA \\
  $ ^{~7}$ High Energy Physics Division, Argonne National Laboratory, Argonne, IL, USA \\
  $ ^{~8}$ Department of Physics, University of Wisconsin, Madison, WI, USA \\
  $ ^{~9}$ Lawrence Berkeley National Laboratory, Berkeley, CA, USA \\
  $ ^{10}$ Institut f\"ur Theoretische Physik, Universit\"at Wien, Wien, Austria \\
  $ ^{11}$ Stanford Linear Accelerator Center, Stanford, CA, USA \\
  $ ^{12}$ Laboratoire de Physique Theorique, Annecy-le-Vieux, France \\
  $ ^{13}$ Department of Physics and Astronomy, Michigan State University,
           East Lansing, MI, USA \\
  $ ^{14}$ Royal Holloway University of London, Egham, Surrey, UK \\
  $ ^{15}$ Skobeltsyn Institute of Nuclear Physics, MSU, Moscow, Russia \\
  $ ^{16}$ Fermi National Accelerator Laboratory, Batavia, IL, USA \\
  $ ^{17}$ Department of Physics, Chonbuk National University, Chonju, Korea \\
  $ ^{18}$ PH Department, CERN, Geneva, Switzerland \\
  $ ^{19}$ Physikalisches Institut, Universit\"at Freiburg, Freiburg, Germany \\
  $ ^{20}$ Physics Department, Universidad Catolica de Chile, Santiago, Chile \\
  $ ^{21}$ LAL, Universit\'{e} de Paris-Sud, IN2P3-CNRS, Orsay, France \\
  $ ^{22}$ University of Delhi, Delhi, India \\
  $ ^{23}$ Institut f\"ur Hochenergiephysik, \"Osterreichische Akademie der
           Wissenschaften, Wien, Austria \\
  $ ^{24}$ Instituto de F\'isica, UNAM, M\'exico, Mexico \\
  $ ^{25}$ Institut f\"ur Theoretische Physik und Astrophysik, 
           Universit\"at W\"urzburg, W\"urzburg, Germany \\
  $ ^{26}$ Institut f\"ur Theoretische Physik, Universit\"at Z\"urich,
           Z\"urich, Switzerland \\
  $ ^{27}$ Max-Planck-Institut f\"ur Physik, M\"unchen, Germany \\
  $ ^{28}$ Centre for High Energy Physics, Indian Institute of Science,
           Bangalore, India \\ 
  $ ^{29}$ Department of Theoretical Physics, Aristotle University of Thessaloniki,
           Thessaloniki, Greece \\
  $ ^{30}$ Facultat de F\'{i}sica, Universitat de Barcelona, Barcelona, Spain \\
  $ ^{31}$ Department of Physics, University of California, Davis, CA, USA \\
  $ ^{32}$ Institute of Theoretical Physics, University of Tokushima, 
           Tokushima, Japan \\
  $ ^{33}$ Santa Cruz Institute for Particle Physics, University of California, 
           Santa Cruz, CA, USA \\
  $ ^{34}$ Theory Division, KEK, Tsukuba, Japan \\
  $ ^{35}$ Department of Modern Physics,
           University of Science and Technology of China, Hefei, China \\
  $ ^{36}$ Center for High Energy Physics and Institute of Modern Physics,
           Tsinghua University, Beijing, China \\
  $ ^{37}$ High Energy Physics, Uppsala University, Uppsala, Sweden \\
  $ ^{38}$ Department of Physics, Tokyo Gakugei University, Tokyo, Jpan \\
  $ ^{39}$ Instituto de F\'{\i}sica Corpuscular, CSIC, Val\`encia, Spain \\
  $ ^{40}$ Department of Physics, National Taiwan University, Taipei, Taiwan \\ 
  $ ^{41}$ Department of Mathematical Sciences, University of Liverpool,
           Liverpool, UK \\
  $ ^{42}$ Institute of Theoretical Physics, Warsaw Univerity, Warsaw, Poland \\
  $ ^{43}$ MCTP, University of Michigan, Ann Arbor, MI, USA \\
  $ ^{44}$ School of Physics, Seoul National University, Seoul, Korea \\
  $ ^{45}$ Department of Physics, Yonsei University, Seoul, Korea \\
  $ ^{46}$ School of Physics and Astronomy, University of Southampton,
           Southampton, UK \\
  $ ^{47}$ Physikalisches Institut der Universit\"at Bonn, Bonn, Germany \\
  $ ^{48}$ Laboratoire de Physique Subatomique et de Cosmologie,
           Universit\'{e} Grenoble I, Grenoble, France \\
  $ ^{49}$ LPTA, Universit\'e Montpellier II, CNRS-IN2P3, Montpellier, France \\
  $ ^{50}$ Institut f\"ur Theoretische Physik, RWTH Aachen, Aachen, Germany \\
  $ ^{51}$ Laboratoire de Physique des Particules, Annecy-le-Vieux, France \\
  $ ^{52}$ Department of Physics and Astronomy, University of Pennsylvania, 
           Philadelphia, PA, USA \\
  $ ^{53}$ Department of Physics, Carleton University, Ottawa, ON, Canada \\ 
  $ ^{54}$ I.\ Physikalisches Institut der RWTH Aachen, Aachen, Germany \\
  $ ^{55}$ Department of Physics, University of Florida, Gainesville, FL, USA \\
  $ ^{56}$ Department of Physics and Astronomy, University of Glasgow, Glasgow, UK \\
  $ ^{57}$ ICEPP, University of Tokyo, Tokyo, Japan \\
  $ ^{58}$ IPN Universit\'{e} Lyon, IN2P3-CNRS, Lyon, France \\ 
  $ ^{59}$ Harish-Chandra Research Institute, Allahabad, India \\
  $ ^{60}$ University of Colorado, Boulder, CO, USA \\
  $ ^{61}$ YITP, Kyoto Universty, Kyoto, Japan \\
  $ ^{62}$ Deutsches Elektronen-Synchrotron DESY, Zeuthen, Germany \\
  $ ^{63}$ William I. Fine Theoretical Physics Institute, University of Minnesota, 
           Minneapolis, MN, USA \\
  $ ^{64}$ INFN, Sezione di Pavia, Pavia, Italy \\
  $ ^{65}$ Department of Physics and Astronomy, University of Rochester,
           Rochester, NY, USA \\
  $ ^{66}$ IPPP, University of Durham, Durham, UK \\
  $ ^{67}$ Tata Institute of Fundamental Research, Mumbai, India \\
  $ ^{68}$ Paul Scherrer Institut, Villigen, Switzerland \\
  $ ^{69}$ Institut f\"ur Experimentalphysik, Universit\"at Hamburg, 
           Hamburg, Germany \\
  $ ^{70}$ Department of Physics, Chung-Ang University, Seoul, Korea \\
  $ ^{71}$ Department of Physics and Astronomy, University of Sheffield,
           Sheffield, UK \\
  $ ^{72}$ Enrico Fermi Institute, University of Chicago, Chicago, IL, USA \\
  $ ^{73}$ Institute of High Energy Physics, Chinese Academy of Sciences, 
           Beijing, China \\
  $ ^{74}$ Department of Physics, Tohoku University, Sendai, Japan \\ 
  $ ^{75}$ ITP, School of Physics, Peking University, Beijing, China \\
%
%
\\ 
  $ ^{~a}$ Supported in part by US DOE, Div. of HEP, 
           contract W-31-109-ENG-38 \\
  $ ^{~b}$ Supported in part by UNAM grant PAPIIT-IN116202 and Conacyt grant 42026-F \\
  $ ^{~c}$ Heisenberg Fellow \\
  $ ^{~d}$ Supported by grant KBN 2~P03B~040~24 \\
  $ ^{~e}$ Supported by a MCyT Ramon y Cajal contract \\
} 



\date{\today}

\abstract{ 
  High-precision analyses of supersymmetry parameters    
  aim at reconstructing the fundamental supersymmetric theory
  and its breaking mechanism. 
  A well defined theoretical framework is needed when  
  higher-order corrections are included.
  We propose such a scheme, Supersymmetry Parameter Analysis SPA, 
  based on a consistent set of conventions
  and input parameters. A repository for computer programs is provided
  which connect parameters in different schemes and   relate
  the Lagrangian parameters to physical observables at LHC and high energy
  $e^+e^-$ linear collider experiments, i.e., masses, mixings, decay widths
  and production cross sections for supersymmetric particles. In
  addition, programs for calculating high-precision low energy observables, 
  the density of cold dark matter (CDM)
  in the universe as well as the cross sections for 
  CDM search experiments are included. 
  The SPA scheme still requires extended
  efforts on both the theoretical and experimental side before data 
  can be evaluated in the future at the level of the desired precision. 
  We take here an initial step of   
  testing the SPA scheme by applying the techniques involved to a
  specific supersymmetry reference point.
  \vspace*{-65.5\baselineskip}
 }

\thispagestyle{empty}

\maketitle
\newpage\noindent%
\setcounter{page}{1}%
\markboth{Supersymmetry Parameter Analysis: SPA Convention and Project}
         {J.A.~Aguilar-Saavedra et al.}
\nobreak

\vspace*{-3\baselineskip}
\section{INTRODUCTION}
At future colliders, experiments can be performed in the supersymmetric
particle sector \cite{SUSY1,SUSY2,SUSY3,SUSY4}, 
if realized in Nature, with very high precision. 
While the Large Ha\-dron Collider LHC can provide us with
a set of well-determined observables~\cite{R2,R2A}, 
in particular masses of colored
particles and precise mass differences of various particle combinations,
experiments at the International $e^+e^-$ Linear Collider ILC \cite{R3,R3A,R3B} 
offer high-precision
determination of the non-colored supersymmetry sector. Combining the
information from LHC on the generally heavy colored particles
with the information from ILC on the generally lighter non-colored
particle sector (and later from the Compact Linear Collider CLIC \cite{clic} on heavier states) 
will generate a comprehensive 
high-precision picture of supersymmetry at the
TeV scale \cite{Allanach:2004ed}. 
Such an analysis can be performed independently of
specific model assumptions and for any supersymmetric scenario that
can be tested in laboratory experiments. It may subsequently serve
as a solid base for the reconstruction of the fundamental supersymmetric
theory at a high scale, potentially close to the Planck scale, 
and for the analysis of the microscopic mechanism of
supersymmetry breaking \cite{Blair:2000gy,Kan}.   

The analyses will be based on experimental accuracies expected at the
percent down to the per-mil level \cite{R3B,R6A}.
These experimental accuracies must
be matched on the theoretical side. This 
demands a well-defined framework for the calculational schemes in
perturbation theory as well as for the input parameters. The proposed 
Supersymmetry Parameter Analysis Convention (SPA) [Sect.2] 
provides a clear base for calculating masses, mixings, decay widths and 
production cross sections. They will serve to extract the fundamental
supersymmetric Lagrangian parameters and the super\-symmetry-breaking
parameters from future data.  
In addition, the renormalization group
techniques must be developed for all the scenarios 
to determine the high-scale parameters of the
supersymmetric theory and its microscopic breaking mechanism. 

By constructing such a coherent and
unified basis, the comparison between results from different calculations
can be streamlined, eliminating ambiguous procedures and reducing
confusion to a minimum when cross-checking results.

\begin{table*}
\begin{center}
\framebox{ \framebox{ \parbox{14.5cm}{
\vspace*{4mm}
\begin{minipage}{13.9cm}
\centerline{\underline{\it SPA CONVENTION}}
\vspace*{2mm}
\begin{itemize}
   \item  The masses of the SUSY particles and Higgs bosons are defined as pole
        masses.
   \item All SUSY Lagrangian parameters, mass parameters and couplings, 
        including $\tan\beta$, are given in the $\drbar$ scheme and defined 
        at the scale $\tilde M =$ 1 TeV.
   \item Gaugino/higgsino and scalar mass matrices, rotation matrices and the
        corresponding angles are defined in the $\drbar$ scheme at $\tilde M$,
        except for the Higgs system in which the mixing matrix is defined in 
        the on-shell scheme, the momentum scale chosen as the light Higgs 
        mass.
   \item The Standard Model input parameters of the gauge sector are chosen as
        $G_F$, $\alpha$, $M_Z$ and $\alpha_s^{\msbar}(M_Z)$. 
        All lepton masses are defined on-shell. 
        The $t$ quark mass is defined on-shell; the $b,\, c$
        quark masses are introduced in $\msbar$ at the scale of the
        masses themselves while taken at a renormalization scale of 2 GeV for
        the light $u,\, d,\, s$ quarks.
   \item Decay widths/branching ratios and production cross sections are
        calculated for the set of parameters specified above.
\end{itemize}
\vskip4mm
\end{minipage}
} } 
} 
\end{center}
\caption{\it Definition of the supersymmetry parameter convention SPA}
\label{tab:SPAconv}
\end{table*}

\newcommand{\half}{\ensuremath{\textstyle \frac12}}
\newcommand{\mrm}[1]{\mathrm{#1}}

A program repository [Sect.3] 
has therefore been built in which a series of programs
has been collected that will be expanded continuously in the future. 
The programs relate parameters defined in different schemes
with each other, e.g. pole masses with $\drbar$ masses, and they calculate
decay widths and cross sections from the basic Lagrangian parameters. An
additional set of programs predicts the values of high-precision
low-energy observables of Standard Model (SM) particles in supersymmetric theories. 
The program repository also
includes global fit programs by which the entire set of Lagrangian
parameters, incorporating higher-order corrections, can be extracted from
the experimental observables. In addition, the solutions of 
the renormalization group
equations are included by which extrapolations from the laboratory
energies to the Grand Unification (GUT) and Planck scales can be performed and {\it vice versa}.
Another category contains programs which relate the supersymmetry (SUSY) parameters with the
predictions of cold dark matter in the universe and the corresponding
cross sections for search experiments of cold dark matter (CDM) particles. 

It is strongly recommended that  the programs available in the
repository adopt the structure of Ref.~\cite{Chung:2003fi}
for the Lagrangian, including flavor mixing and CP phases, and follow 
the generally accepted Supersymmetry Les Houches Accord,
SLHA, for communication between different programs \cite{Skands:2003cj}.
For  definiteness, we reproduce from \cite{Skands:2003cj}
the superpotential (omitting $R$-parity violating terms), 
in terms of superfields,
\begin{eqnarray}
W &=& \epsilon_{ab} \left [ 
      (Y_E)_{ij} \hat H_d^a \hat L_i^b    \hat{\bar E}_j 
    + (Y_D)_{ij} \hat H_d^a \hat Q_i^{b}  \hat{\bar D}_{j}  \right .  
\nonumber\\
  & & \left. \phantom{xx}
    + (Y_U)_{ij} \hat H_u^b \hat Q_i^{a} \hat{\bar U}_{j} 
    - \mu \hat H_d^a \hat H_u^b \right ] \, ,
\label{eq:superpot}
\end{eqnarray}
where the chiral superfields of the Minimal Supersymmetric Standard
Model (MSSM) have the following $SU(3)_C\otimes
SU(2)_L\otimes U(1)_Y$ quantum numbers
\begin{eqnarray*}
&& \hat L:(1,2,-\half),\, \hat{\bar E}:(1,1,1),\, \textstyle
   \hat Q:(3,2,\frac16),\,\hat{\bar U}:(\bar{3},1,-\frac{2}{3}) \\ 
&& \hat{\bar D}:(\bar{3},1,\textstyle \frac13),\, 
   \hat H_d:(1,2,-\half),\, \hat H_u:(1,2,\half)\, .
\label{fields}
\end{eqnarray*}
The indices of the  $SU(2)_L$ fundamental representation are denoted 
by $a,b=1,2$ and the generation indices by $i,j=1,2,3$. Color indices
are everywhere suppressed, since only trivial contractions are involved.
$\epsilon_{ab}$ is the totally antisymmetric tensor, with
$\epsilon_{12}=\epsilon^{12}=1$.

The soft SUSY breaking part is written as
\begin{eqnarray}
&&-{\cal L}_{soft} = \epsilon_{ab} \left [
  (T_E)_{ij} H_d^a  \tilde{L}_{i_L}^{b} \tilde{e}_{j_R}^* +
  (T_D)_{ij} H_d^a  \tilde{Q}_{i_L}^{b}  \tilde{d}_{j_R}^* \right .  
\nonumber\\
&& \phantom{mmmmmmm} \left . +
   (T_U)_{ij}  H_u^b \tilde{Q}_{i_L}^{a} \tilde{u}_{j_R}^*
   \right ] + \mrm{h.c.} 
\nonumber\\
&&~+ m_{H_d}^2 {{H^*_d}_a} {H_d^a} + m_{H_u}^2 {{H^*_u}_a} {H_u^a} 
   - (m_3^2 \epsilon_{ab} H_d^a H_u^b + \mrm{h.c.}) 
\nonumber\\
&&~+ {\tilde{Q}^*}_{i_La} (m_{\tilde Q}^2)_{ij} \tilde{Q}_{j_L}^{a} 
   + {\tilde{L}^*}_{i_La} (m_{\tilde L}^2)_{ij} \tilde{L}_{j_L}^{a}  
\nonumber\\
&&~+ \tilde{u}_{i_R} (m_{\tilde u}^2)_{ij} {\tilde{u}^*}_{j_R} 
   + \tilde{d}_{i_R} (m_{\tilde d}^2)_{ij} {\tilde{d}^*}_{j_R} 
   + \tilde{e}_{i_R} (m_{\tilde e}^2)_{ij} {\tilde{e}^*}_{j_R} 
\nonumber\\
&&~+ \frac{1}{2} \left( M_1 {\tilde b}{\tilde b} 
   + M_2 {\tilde w}^A{\tilde w}^A 
   + M_3 {\tilde g}^X {\tilde g}^X \right) + \mrm{h.c.} \, ,
\end{eqnarray}
where the $H_i$ are the scalar Higgs fields, the
fields with a tilde are the scalar components of the superfield
with the identical capital letter; the bino is denoted as ${\tilde b}$, the
unbroken $SU(2)_L$ gauginos as ${\tilde w}^{A=1,2,3}$, and the gluinos as
${\tilde g}^{X=1\ldots8}$, in 2-component notation.
The $T$ matrices will be decomposed as 
$T_{ij} = A_{ij} Y_{ij}$, 
where $Y$ are the Yukawa matrices and $A$ the soft supersymmetry breaking
trilinear couplings.

Much work on both the theoretical and the experimental side is still
needed before data could be evaluated in the future at the desired level
of accuracy. These tasks of the SPA Project will be defined   
in detail in Sect.4. 

In Sect.5 we introduce the SUSY reference point SPS1a$'$ as a general 
setup for testing these tools in practice. This reference point is defined 
at a characteristic scale of 1 TeV in the Minimal Supersymmetric Standard Model
with roots in minimal supergravity (mSUGRA). The point is
a derivative of the Snowmass point SPS1a \cite{Allanach:2002nj};
its parameters are identical
except for a small shift of the scalar mass parameter 
and a change of the trilinear coupling to comply with the measured dark matter  
density~\cite{WMAP}.
Note, that the SPS1a$'$ parameters are compatible with all the available high-
and low-energy data.
The parameters are close to point B$'$ of
Ref.~\cite{Battaglia:2003ab}. 
The masses are fairly light
so that stringent tests of all aspects in the program can be performed
for LHC and ILC experiments. 
The final target are predictions on the
accuracies of the fundamental supersymmetry parameters that can be
expected from a common set of information when LHC and ILC 
experiments are analyzed cohe\-rently. 

Additional benchmark points within and beyond mSUGRA, 
representing characteristics of different
scenarios, should complement the specific choice of~SPS1a$'$.

\section{SPA CONVENTION}

Extending the experience collected in analyzing Standard Model parameters
at the former $e^+e^-$ colliders LEP and
SLC, we propose the set of conventions defined in Table~\ref{tab:SPAconv}.
These conventions conform with the general SLHA scheme\cite{Skands:2003cj}
but they are more specific in several points. 

Though largely accepted as standard,
some of the definitions proposed in this SPA Convention should be explained 
in a few comments. 

For the SUSY Lagrangian parameters the $\drbar$  sche\-me
\cite{Siegel:1979wq,Jack:1994rk} is most useful.  It
is based on regularization by dimensional reduction together with modified
minimal subtraction. This scheme is designed
to preserve supersymmetry by maintaining the number of degrees of
freedom of all fields in $D$ dimensions,
and it is technically very convenient.
The
$\beta$-functions for SUSY parameters in this scheme are known up to 3-loop
order \cite{JJK}. It has recently been shown
\cite{Stockinger:2005gx} that  
inconsistencies of the original sche\-me \cite{Siegel:1980qs} 
can be overcome and that
the $\drbar$ scheme can be formulated in a mathematically consistent way.
The ambiguities associated with the treatment of the
Levi-Civita tensor can be parameterized as renormalization scheme dependence
as was argued in \cite{JJ}.  
Checks by explicit evaluation of the supersymmetric
Slavnov-Taylor identities  at the one-loop level
have shown that the $\drbar$ method generates the
correct counter terms \cite{Hollik:1999xh}.
[We will use
the version of the $\drbar$ scheme as given in \cite{Jack:1994rk}, 
there referred to as $\drbar'$ scheme.]
To make use of the highly developed
infrastructure for proton colliders, which is based on the $\msbar$ 
factorization scheme \cite{Beenakker:1996ch},  
a dictionary is given in Sect.3.2 for
the translation between the $\drbar$ and $\msbar$ schemes, 
as well as the on-shell renormalization schemes. 

The SUSY scale is chosen $\tilde M = 1$~TeV to avoid large 
threshold corrections in running the mass parameters by renormalization group
techniques from the high scale down to the low scale. 
Fixing the scale $\tilde{M}$ independent of parameters
within the supersymmetry scenarios is preferable over choices
relating to specific parameters, such as squark masses, that can be fixed only
at the very end.  By definition, this point can also be used to characterize
uniquely multiple-scale approaches.

Mixing parameters, in particular $\tan\beta$,
could have been introduced in different ways \cite{Freitas:2002um}; 
however, choosing the $\drbar$ definitions proposed above has proven very
convenient in practical calculations. 


The masses of Higgs bosons~\cite{HiggsRev}, in the MSSM of the charged $H^\pm$, of the
neutral {\it CP}-odd $A$, and of the two {\it CP}-even $h,H$ particles, 
are understood as pole masses, $M_{H^\pm,A,H,h}$.
For given $M_A$, the pole masses $M_{H,h}$ of the {\it CP}-even Higgs bosons
are obtained as poles $q^2=M_{H,h}^2$
of the dressed propagator matrix, 
\begin{eqnarray*}
\small
\Delta_{\rm Hh}(q^2) &= 
   \scriptstyle 
   \left( \begin{array}{cc} 
   q^2 \!-\!  m_H^2 \!+\! \Sigma_{HH}(q^2) &  \Sigma_{hH}(q^2) \\[0.1cm]
     \Sigma_{hH}(q^2) & q^2 \!-\!  m_h^2 \!+\! \Sigma_{hh}(q^2) 
\end{array} \right)^{-1}
\\
\end{eqnarray*}
involving  the tree-level masses $m_{H,h}$ and 
the diagonal and  non-diagonal 
on-shell-renormalized self-energies $\Sigma$.
In the on-shell scheme, the input parameters are renormalized
on-shell quantities, in particular the $A$-boson mass, with
accordingly defined counter terms. 

Owing to the momentum dependence of the self-energies,
there is no unique mixing angle ($\alpha$) 
for the neutral {\it CP}-even Higgs system beyond the tree level,
and the SPA choice can be understood as a convention
for an ``improved Born approximation''.
A convenient choice for $q^2$ in the self-energies 
which minimizes the difference of such an approximation with respect to 
calculations involving the proper self-energies in physical matrix elements,
is given by $q^2=M_h^2$. 

The physical on-shell masses are introduced in the decay widths and
production cross sections such that the phase space is treated
in the observables closest to experimental on-shell kinematics. This 
applies to the heavy particles while the masses of the light particles 
can generally be neglected in high energy processes. 

In the chargino/neutralino sector the number of observable masses exceeds
the number of free parameters in the system,
gaugino/higgsino mass parameters and $\tan\beta$. The
most convenient set of input chargino/neu\-tralino masses is
dictated by experiment [the three lowest mass states in this
sector, for example] while the additional masses are subsequently 
predicted uniquely. Similar procedures 
need to be followed in the sfermion sector.

\section{PROGRAM BASE } 
\label{prog-base}
\subsection{PROGRAM CATEGORIES}

The computational tasks that are involved in the SPA Project can be
broken down to several categories.  Each of the codes that will be
collected in the SPA program repository is included 
in one or more of these categories.
It is understood that in each case the theoretical state-of-the-art
precision is implemented. For communication between codes SLHA
\cite{Skands:2003cj} 
is strongly recommended, which is extended in a suitable way where appropriate.

\begin{itemize}
  \item[1)] \underline{Scheme translation tools:}

    The communication between codes that employ different calculational
    schemes requires a set of translation rules.  In the SPA program
    repository we there\-fore collect tools that implement, in particular, the
    definitions and relations between on-shell, $\drbar$ and
    $\msbar$ parameters in the Lagrangian as listed in Sect. 3.2 below.

  \item[2)] \underline{Spectrum calculators:}

    This category includes codes of 
    the transition from the Lagrangian parameters to a
    basis of physical particle masses and the related mixing matrices. 
    This task mainly consists of deriving the on-shell particle masses
    (including higher-order corrections)   
    and of diagonalizing the mixing matrices in a consistent
    scheme, making use of the abovementioned tools as needed.
    
  \item[3)] \underline{Calculation of other observables:}
    \begin{itemize}
    \item[3A)] Decay tables: \\
      compute the experimentally measurable widths and branching
      fractions.
      
    \item[3B)] Cross sections: \\
      calculate SUSY cross sections and distributions for LHC and ILC.

    \item[3C)] Low-energy observables: \\
      compute the values of those low-energy, high-precision observables 
      [{\it e.g.},
      $b\to s \gamma$, $B_s \to \mu\mu$, $g_\mu-2$] that are sensitive to SUSY
      effects. 

    \item[3D)] Cosmological and astrophysical aspects: \\
      this category of programs covers the derivation of cold dark matter
      (CDM) relic density in the universe, 
      cross sections for CDM particle search\-es, astrophysical
      cross sections, {etc.} in the SUSY context.
    \end{itemize}

  \item[4)] \underline{Event generators:}

    Programs that generate event samples for SUSY and background
    processes in realistic collider environments.

  \item[5)] \underline{Analysis programs:}
    
    These codes make use of some or all of the above to extract the
    Lagrangian parameters from experimental data by means of global
    analyses.    
    
  \item[6)] \underline{RGE programs:}
    
    By solving the renormalization-group equations, the programs connect the
    values of the parameters of the low-energy effective Lagrangian to those
    at the high-scale where the model is supposed to match to a more
    fundamental theory.  
    High-scale constraints are implemented on the basis of
    well-defined theoretical assumptions: gauge coupling unification,
    \linebreak[4]
    mSUGRA, GMSB, AMSB scenarios, {etc}.
    
  \item[7)] \underline{Auxiliary programs and libraries:}
    
    Structure functions, beamstrahlung, numerical \linebreak[4]
    methods, SM
    backgrounds, {etc}.
    
  \end{itemize}

  This is an open system and the responsibility for all these programs 
  remains with the authors. SPA
  provides the translation tables and the links to the computer codes on
  the web-page
  \begin{center}
    {\ttfamily{  http://spa.desy.de/spa/  }}
  \end{center}
  Conveners responsible for specific tasks of the SPA Project will be listed
  on this web-page; the information will be routinely updated to reflect the
  momentary state of the project at any time.

\subsection{SCHEME TRANSLATION} 

This subsection presents a few characteristic examples 
of relations between on-shell observables and $\drbar$, $\msbar$ quantities
at the electroweak scale $M_Z$ and the SUSY scale $\tilde{M}$. For brevity,
here only the approximate one-loop results are given
\cite{Martin:1993yx}; 
it is understood that the codes in the program repository include the
most up-to-date higher-loop results.  
\begin{itemize}
  \item[(a)] \underline{Couplings:}   \\[2mm]
    $\bullet$ {\it gauge couplings}:
    \begin{equation}
      \small
      g_i^{\msbar} = g_i^{\drbar} \left( 1-\frac{(g_i^{\drbar})^2}{96\pi^2} C_i \right)
    \end{equation}      
    $\bullet$ {\it Yukawa couplings between the gaugino $\lambda_i$, the chiral fermion $\psi_k$
      and the scalar $\phi_k$}:
    \begin{equation}
      \small
      \hat g^{\msbar}_{ik} = g_i^{\drbar}
               \left ( 1 + \frac{(g_i^{\drbar})^2}{32\pi^2} C_i
                         - \sum_{l=1}^3 \frac{(g_l^{\drbar})^2}{32\pi^2} C_l^{r_k} \right )
    \end{equation}
    $\bullet$ {\it Yukawa couplings between the scalar $\phi_i$ and the
      two chiral fermions $\psi_j$ and $\psi_k$}:
    \begin{equation}
      \small
      Y_{ijk}^{\msbar} = Y_{ijk}^{\drbar} \bigg(1+\sum_{l=1}^3
      \frac{(g_l^{\drbar})^2}{32\pi^2}\big[ C_l^{r_j} - 2C_l^{r_i} + C_l^{r_k} \big]\bigg)
    \end{equation}
    $\bullet$ {\it trilinear scalar couplings}:  \\[2mm]
    These couplings do not differ in the two schemes.\\  

    $C_i$ and $C_i^r$ are the quadratic Casimir invariants 
    of the adjoint representation and the matter representation $r$
    of the gauge group $G_i$, respectively. They are given by
    $C_i = [3,2,0]$ for $[SU(3), SU(2), U(1)]$ 
    and $C_i^r = [4/3, 3/4, 3/5 \times Y_r^2]$ for the fundamental
    representations 
    of $SU(3), SU(2)$, and the $U(1)$ hypercharge $Y_r$.  \\
  \item[(b)] \underline{SUSY $\drbar$, $\msbar$ and pole masses:} \\[2mm]
    $\bullet$ {\it gaugino mass parameters} 
    \begin{eqnarray}
      M_i^{\msbar} &=& M_i^{\drbar} \left( 1+\frac{(g_i^{\drbar})^2}{16\pi^2} C_i \right)
    \end{eqnarray}
    $\bullet$ {\it higgsino mass parameter}:
    \begin{eqnarray}
      \mu^{\msbar} = \mu^{\drbar}\left(1+\sum_{l=1}^2
        \frac{(g_l^{\drbar})^2}{16\pi^2} C^H_l \right)         
    \end{eqnarray}
    $C_l^H$ denoting the $SU(2) \;{\rm and}\; U(1)$ Casimir invariants 
    of the Higgs fields.  \\
    
    $\bullet$ {\it sfermion mass parameters}: \\[2mm] 
    These parameters do not differ in the $\drbar$ and $\msbar$ schemes.\\[\baselineskip]

    $\bullet$ {\it fermion pole masses}:  \\[2mm]
    The pole masses can be written schematically as
    \begin{eqnarray}
      m_{i,\, \rm pole} &=& M_i^\drbar - {\rm Re}\,\Sigma\,(\fmslash q=m_{i,\,\rm pole})  
    \end{eqnarray}
    where $\Sigma$ denotes the fermion self-energy
    renormalized according to the $\drbar$-scheme at the scale $\tilde{M}$.
    As an explicit example we note the one-loop relation between
    the SU(3) gaugino mass parameter
    $M_3(\tilde{M})^\drbar$ and the gluino pole mass $m_{\tilde{g}}$ 
    [without sfermion mixing] at the one-loop order:
    \begin{eqnarray}
      m_{\tilde g}  &=&  M_3^{\drbar}(\tilde M)       \\ 
      & & + \ \frac{\alpha_s^\drbar(\tilde M)}{4 \pi}
      \bigg[ m_{\tilde g} \bigg(15 + 9 \ln \frac{\tilde M^2}{m_{\tilde g}^2}\bigg) 
      \nonumber\\
      & & \hspace*{1.9cm}
      + \ \sum_q \sum_{i=1}^2 m_{\tilde g} 
      B_1\left(m^2_{\tilde g}, m^2_q, m^2_{\tilde q_i}\right)  \bigg]
      \nonumber
    \end{eqnarray}
    where $B_1$
    is the finite part of one of the one-loop two-point
    functions  
    at the scale in the $\drbar$ scheme $\tilde M$
    (and analogously $A_0,B_0$ to be used later), cf.\ Ref.~\cite{Denner:1991kt}.   \\
    
    $\bullet$ {\it scalar pole masses}:  \\[2mm]
    A similar relation holds for the squared scalar masses
    \begin{eqnarray}
      m^2_{i,\,{\rm pole}}=M^{2,\,\drbar}_i - \Sigma(q^2=m^2_{i,\, \rm pole})
    \end{eqnarray}
    The one-loop QCD corrections for the left squarks of the 
    first two generations in the limit of vanishing quark masses
    may serve as a simple example:
    \begin{eqnarray}
      m^2_{\tilde q} &=& M^{2,\,\drbar}_{\tilde Q}(\tilde M) 
      \\ & &
       - \ \frac{2 \alpha_s^\drbar(\tilde M)}{3 \pi}
      \bigg[(m^2_{\tilde q} - m^2_{\tilde g}) 
               B_0(m^2_{\tilde q},m^2_{\tilde g},0) 
      \nonumber\\ 
      && \hspace*{3mm}
      - \ 2 m^2_{\tilde q} B_0(m^2_{\tilde q},m^2_{\tilde q},0)
      +  A_0( m^2_{\tilde q}) - A_0( m^2_{\tilde g})
      \bigg]
      \nonumber
    \end{eqnarray}
    \\
   
  \item[(c)] \underline{SM parameters}:\\[2mm]
    The following paragraphs summarize the SM input values for the analysis.
    Only approximate formulae are presented for brevity,
    while the complete set of relations is available on the program
    repository. 
  
    In a few 
    cases the evolution from the scale $M_Z$ to $\tilde M$ is carried out
    by means of RGEs instead of fixed-order perturbation theory because
    they have proven, presently, more accurate; this may change 
    once the necessary multi-loop calculations will be completed. \\
    
    $\bullet$ 
    $\alpha$, $\alpha^{\drbar}(M_Z)$, $\alpha^{\drbar}_{1,2}(\tilde M)$:
    \begin{eqnarray}
      \alpha^{\drbar}(M_Z) &=& \frac{\alpha}{1 - \Delta \alpha_{\rm SM}
        -  \Delta \alpha_{\rm SUSY}}  \\[.5em]
      \Delta \alpha_{\rm SUSY} &=&
       - {\alpha\over 6\pi}\,\Biggl[ \ln \frac{m_{H^+}}{M_Z} 
      \ +\ 4 \sum_{i=1}^2\ln {m_{\tilde\chi_i^+}\over M_Z}
      \nonumber\\ 
      && \hspace*{10mm} +
      \ \sum_f \sum_{i=1}^2 N_c Q_f^2 \ln {m_{\tilde f_i}\over M_Z}
      \ \Biggr]
      \nonumber
    \end{eqnarray}
    $\Delta \alpha_{\rm SM}$ summarizes the SM contributions from the leptons,
    quarks and the $W$-boson. In the SUSY contributions, $\Delta \alpha_{\rm SUSY}$, $f$ sums over
    all charged sfermions,  $N_c$ is the color factor and $Q_f$ the
    (s)ferm\-ion charge. 
    \begin{eqnarray}
      \alpha_1^\drbar(\tilde{M}) &=& 
      \frac{\displaystyle \frac{ \alpha^\drbar(M_Z) }{ \cos^2 \theta^{\drbar}(M_Z)}}
      {\displaystyle 1 + \frac{1}{4 \pi} \frac{\alpha^\drbar(M_Z)}{\cos^2 \theta^{\drbar}(M_Z)}
                         \ln \frac{M^2_Z}{\tilde M^2}}
      \\[.5em]
      \alpha_2^\drbar(\tilde{M}) &=& 
      \frac{\displaystyle \frac{ \alpha^\drbar(M_Z) }{ \sin^2 \theta^{\drbar}(M_Z)}}
      {\displaystyle 1 + \frac{1}{4 \pi} \frac{\alpha^\drbar(M_Z)}{\sin^2 \theta^{\drbar}(M_Z)}
                         \ln \frac{M^2_Z}{\tilde M^2}} 
    \end{eqnarray}

    $\bullet$ {\it $\sin^2\theta^\drbar$ at $M_Z$ and at $\tilde{M}$}: \\[2mm]
    The electroweak mixing parameter $\sin^2\theta^\drbar(M_Z)$ is given by
    \begin{eqnarray}
      \sin^2\theta^\drbar(M_Z) \left[1 -\sin^2\theta^\drbar(M_Z)\right] &&
      \phantom{xxxxx} 
      \nonumber\\    
      & & \hspace*{-35mm} = \ 
      \frac{ \pi \alpha^\drbar(M_Z)}
              {\sqrt{2} M^2_Z G_F (1 - \Delta \hat r)}
    \end{eqnarray}
    where the contributions from loops of SM and SUSY particles 
    are denoted by $\Delta \hat r$ \cite{Deltar,deltarSUSY}. 
    At the scale $\tilde{M}$ the
    electroweak mixing parameter can be calculated subsequently from
    \begin{eqnarray}
      \tan^2 \theta^{\drbar}(\tilde{M}) = \alpha_1^{\drbar}(\tilde{M})
      / \alpha_2^{\drbar}(\tilde{M})
    \end{eqnarray}
    by making use of the couplings $\alpha_i^{\drbar}(\tilde{M})$ 
    given in the preceeding paragraph. \\
    
    $\bullet$ {\it $\sin^2\theta^{\drbar}$ and $\sin^2\theta_{\rm eff}$
      at $M_Z$:}\\[2mm]
    The electroweak mixing angle in the effective leptonic (electronic)
    vertex of the $Z$ boson is defined as
    \begin{eqnarray}
      \sin^2\theta_{\rm eff} \equiv 
      \sin^2\theta_{\rm eff}^{\rm (e)}(M_Z) =
      \frac{1}{4} \left( 1- {\rm Re}\, \frac{g_V^{\rm e}}{g_A^{\rm e}}
      \right) 
    \end{eqnarray}
    in terms of the effective vector and axial vector couplings 
    $g_{V,A}^{\rm e}$ of the $Z$ to electrons.
    The relation to 
    $\sin^2\theta^\drbar(M_Z)$ is given by (at one-loop order)
    \begin{eqnarray}
      \sin^2\theta^\drbar(M_Z) & = & \sin^2\theta_{\rm eff} 
      \\
      & & \hspace{-10mm} + \ \sin 2\;\theta_{\rm eff}\;
      \frac{\Pi_{\gamma Z}(M_Z^2) +{\Pi_{\gamma Z}(0)}}{2\;M_Z^2} - f^{\rm e}
      \, ,  
      \nonumber
    \end{eqnarray}
    involving the photon--$Z$ non-diagonal self-energy \-
    \mbox{$\Pi_{\gamma Z}(q^2)$} 
    and the non-universal electron--$Z$ vertex correction form factors
    $f_{V,A}^{\rm e}(q^2)$,
    \begin{eqnarray}
      f^{\rm e} = {\textstyle\frac{1}{2}}\, f_V^{\rm e}(M_Z^2) - 
      ({\textstyle\frac{1}{2}} - 2\, \sin^2\theta_{\rm eff} )\, f_A^{\rm e}(M_Z^2),\
    \end{eqnarray}
    with all the loop quantities renormalized in the $\drbar$ scheme 
    at the scale $M_Z$.
    For explicit expressions see~\cite{Deltar,deltarSUSY}. \\
    
    $\bullet$ {\it $\alpha_s^\drbar$ at $M_Z$ and $\tilde M$, related
      to $\alpha_s^\msbar(M_Z)$}: 
    \begin{eqnarray}
      \alpha_s^\drbar(M_Z) &=& {\alpha^{\msbar}_s(M_Z)\over 1-\Delta\alpha_s} 
     \\[.5em]
      \Delta\alpha_s &=& {\alpha_s(M_Z)\over2\pi}\ \biggl[ {1\over2}\ -
      \ {2\over3}\ln {m_t\over M_Z}
      \nonumber\\ 
      && \hspace*{6mm} -\ 2\ln {m_{\tilde g}\over M_Z} \ 
         -\ {1\over6}\sum_{\tilde q}\sum_{i=1}^2\ln {m_{\tilde q_i}\over M_Z}
      \biggr]  
      \nonumber\\[.5em]
      \alpha_s^\drbar(\tilde M) &=& 
      \frac{ \alpha_s^\drbar(M_Z) }
      {1 - \frac{3}{4 \pi} \alpha_s^\drbar(M_Z) \ln \frac{M^2_Z}{\tilde M^2}} 
    \end{eqnarray}
   
    $\bullet$ {\it $W,\, Z$ bosons, pole and $\drbar$ masses}:\\[2mm]
    The pole masses $M_V$ ($V=W,Z$) and the $\drbar$ masses at $M_Z$ 
    are related by
    \begin{eqnarray}
      M^2_V = M^{2,\,\drbar}_V(M_Z) - {\rm Re}\, \Pi^T_{VV}(p^2=M^2_V) 
    \end{eqnarray}
    involving the renormalized transverse vector-boson self-energies 
    in the $\drbar$ scheme at the scale $M_Z$.
    The $Z$ pole mass is a direct input parameter, whereas
    the $W$ pole mass is derived from the relation to the
    low-energy parameters $\alpha$ and Fermi constant $G_F$ according
    to the SPA Convention:
    \begin{eqnarray}
      M_W^2 \left(1-\frac{M_W^2}{M_Z^2}\right) =
      \frac{\pi \alpha}{\sqrt{2} G_F (1-\Delta r)} ,
    \end{eqnarray}
    $\Delta r$ summarizes the loop
    contributions from the SM and SUSY particles as 
    given explicitly in \cite{Deltar,deltarSUSY,deltar2}.
    
    The self-energies at the scale $\tilde M$ can be written symbolically as
    \begin{eqnarray}
      16 \pi^2 \Pi^T_{ZZ} &=& 16 \pi^2 \Pi^T_{ZZ,\,\rm SM+Higgs}  
      \\
      & & - \ \sum_f 4 N^f_c v^2_{fZ,ij}
          \tilde B_{22}(M^2_Z, m^2_{\tilde f_i}, m^2_{\tilde f_j}) 
      \nonumber\\
      & & + \ \sum_{\tilde \chi^0, \tilde \chi^+}
          \big[ f_{ijZ} H(M^2_Z, m_{\tilde \chi_i}, m_{\tilde \chi_j}) 
      \nonumber\\
      & & \hspace{10mm}
       + \ 2 g_{ijZ} B_0(M^2_Z, m_{\tilde \chi_i}, m_{\tilde \chi_j}) \big] 
      \nonumber\\[.5em]
      16 \pi^2 \Pi^T_{WW} &=& 16 \pi^2 \Pi^T_{WW,\,\rm SM+Higgs}  
      \\
      & & - \ \sum_f 2 N^f_c v^2_{fW,ij}
          \tilde B_{22}(M^2_W, m^2_{\tilde f_i}, m^2_{\tilde f'_j}) 
      \nonumber\\
      & & + \ \sum_{i,j}
          \big[ f_{ijW} H(M^2_W, m_{\tilde \chi^0_i}, m_{\tilde \chi^+_j}) 
      \nonumber\\
      & & \hspace{10mm}
        + \ 2 g_{ijW} B_0(M^2_W, m_{\tilde \chi^0_i}, 
                                 m_{\tilde \chi^+_j}) \big] 
      \nonumber
   \end{eqnarray}
   where $v_{fV,ij}$ are the couplings of the gauge boson to sfermions
   and $f_{ijV}$ and $g_{ijV}$ are combinations of left- and right-couplings to
   charginos and neutralinos; $\tilde B_{22}$ and $H$ are combinations of the
   $B_i$ and $A_i$ loop functions. Detailed formulae are given in 
   \cite{Pierce:1996zz}. \\

$\bullet$ {\it charm and bottom running  $\msbar$ mass at $m_{c,b}$ 
and $\drbar$ mass at $M_Z$, cf.~\cite{Baer:2002ek,car}}:
\begin{eqnarray}
  m^{\drbar}_{b,\,\rm SM}(M_Z) &=& m_b^{\msbar}(m_b)
  \left[ \frac{\alpha^\msbar_s(M_Z)}{\alpha^\msbar_s(m_b)}
  \right]^{\frac{12}{23}}   
  \nonumber\\ 
  && \ \times
  \left[ 1 - \frac{\alpha^\drbar_s}{3 \pi}
    - \frac{23 \alpha^{2,\drbar}_s}{72 \pi} \right]
\end{eqnarray}
\begin{eqnarray}
  m_b^{\drbar}(M_Z) &=& \frac{m^{\drbar}_{b,\,\rm SM}(M_Z) 
                           + {\rm Re}\,\Sigma_b'(M_Z)}
           {1- {\Delta m_b(M_Z)}} 
  \\[1mm]
  \Delta m_b(M_Z)
  &=& \frac{2 \alpha_s}{3 \pi} m_{\tilde g} \mu \tan \beta\,
      I(m^2_{\tilde b_1}, m^2_{\tilde b_2}, m_{\tilde g}^2)
  \nonumber\\
  && \hspace*{0mm} + \
  \frac{Y^2_t}{16 \pi^2} A_t \mu \tan \beta \,
  I(m^2_{\tilde t_1}, m^2_{\tilde t_2}, \mu^2)
  \nonumber\\
  && \hspace*{-0mm} - \
  \frac{g^2}{16 \pi^2} M_2 \mu \tan \beta 
  \nonumber\\[1mm] 
  && \hspace*{-0mm} \times \big[ \cos^2 \theta_{\tilde t}\,
  I(m^2_{\tilde t_1}, M^2_2, \mu^2)
  + \frac{1}{2} \{\tilde t \to \tilde b \} 
  \nonumber\\ 
  && 
  \hspace*{-0mm} + \ \{\cos \to \sin; \ \tilde{Q}_1 \to \tilde{Q}_2 \}\big] 
  \nonumber\\[1mm]
  I(a^2,b^2,c^2) &=& \frac{a^2 b^2 \log a^2 / b^2 + {\rm cyclic}} 
  {(a^2-b^2)(b^2-c^2)(a^2-c^2) }
  \nonumber
\end{eqnarray}
with $\Sigma'_b(M_Z) = \Sigma_b(M_Z) -m_b^{\drbar}(M_Z) \Delta m_b(M_Z)$ 
and $\Sigma_b(M_Z)$ being the
self-energy of the bottom quark due to supersymmetric particles and 
heavy SM particles
and $\Delta m_b(M_Z)$ including the 
large finite terms proportional to $\tan\beta$ 
which have been resummed~\cite{car}. 
In the case of the charm quark the additional running 
between $m_c$ and $m_b$ has
to be included. The SUSY contributions are in general small and
no resummation is necessary. The masses are evolved from the scale 
$M_Z$ to $\tilde{M}$
by means of the RGEs for the Yukawa couplings as described below. \\

$\bullet$ {\it top quark pole mass and $\drbar$ mass at $M_Z$}:
\begin{eqnarray}
  m_t^{\drbar}(M_Z) &=& m_t
  \Bigg[ 1 - \frac{5 \alpha^\drbar_s}{3 \pi}-
     \frac{ \alpha^\drbar_s}{\pi} \log\left(\frac{M^2_Z}{m^2_t}\right)
   \nonumber
   \\  &&  \qquad -   
    c_t  \Big(\frac{\alpha^\drbar_s}{\pi}\Big)^2
    - \Sigma  \; \Bigg]         
\end{eqnarray}
where $c_t(M_Z^2/m_t^2)$ is the gluonic two-loop contribution and 
$\Sigma$ accounts for the
electroweak as well as the SUSY contributions.  The mass is evolved
to the scale $\tilde{M}$
by means of the Yukawa RGEs; see next. \\

$\bullet$ {\it Yukawa couplings and running masses of SM particles 
at $\tilde M$}:\\[2mm]
The vacuum expectation values 
$v^{\drbar}_{u}$ and $v^{\drbar}_{d}$ are initially given by:
\begin{eqnarray}
  &&M^2_W(M_Z) = \frac{1}{4} g^{2,\,\drbar}{(M_Z)}\\
  && \hspace*{2cm} \times 
  \left[ {v^{2,\,\drbar}_{u}}(M_Z) + {v^{2,\,\drbar}_{d}}(M_Z) \right] 
  \nonumber\\[.5em]
  &&v^{\drbar}_{u}(M_Z)/v^{\drbar}_{d}(M_Z) = \tan\beta^{\drbar}(M_Z)  
\end{eqnarray}
$\tan\beta^{\drbar}(M_Z)$ must be evolved down from the conventional parameter
$\tan\beta^{\drbar}(\tilde{M})$ by means of RGE. 
From the $\drbar$ masses at $M_Z$ the Yukawa couplings are calculated:
\begin{eqnarray}
  Y^{\drbar}_t(M_Z) &=& {\sqrt{2} m_t^{\drbar}(M_Z)}/{v^{\drbar}_u(M_Z)} \\[.5em]
  Y^{\drbar}_{b,\tau}(M_Z) &=& {\sqrt{2} m_{b,\tau}^{\drbar}(M_Z)}/
  {v^{\drbar}_d(M_Z)} 
\end{eqnarray}
In a second step, 
they are evolved together with the gauge couplings and the vacuum expectation
values to $\tilde M$ via RGEs. At this scale the running SM fermion masses
and gauge boson masses are related to the Lagrangian parameters 
by the usual tree-level relations.
This is, presently, a better approach for the evolution of the Yukawa
couplings than fixed-order perturbation theory.  

\end{itemize}

\subsection{WIDTHS AND CROSS SECTIONS} 

\begin{itemize}
\item[(a)] \underline{Decay widths:}\\[1mm]
The decay widths are defined as inclusive quantities including all radiative 
corrections; the masses of the heavy particles are taken on-shell,
light particle masses are set zero. \\

\item[(b)] \underline{Cross sections for $e^+e^-$ collisions:}\\ 
Cross sections, $\sigma(e^+e^- \to \tilde{\{F\}})$,
for the production of a set of supersymmetric particles/Higgs bosons
$\{\tilde{F}\}$ are defined 
at the experimental level in $e^+e^-$ collisions including  
up-to-date radiative corrections except 
hard $\gamma$ brems\-strahlung
to exclude large contributions from radiative return. 

In general, large  QED-type photonic corrections cannot be 
disentangled from genuine SUSY-specific \linebreak[4] 
parts, and in the comparison 
of theoretical predictions with experimental data
all higher-order terms have to be included. To elucidate the role
of the  specific supersymmetric loop  corrections, 
a reasonable and consistent prescription for 
cut-independent reduced cross sections shall 
therefore be defined.
Since the leading QED terms arising from virtual
and real photon contributions that contain large logarithms
can be identified and isolated,  
the ``reduced'' genuine SUSY cross sections are defined, at the 
theoretical level, by subtracting 
the logarithmic terms $\log 4 \Delta E^2 /s$ 
in the soft-photon energy cut-off $\Delta E$  
 and in $\log s/m^2_f$ from 
non-collinear and collinear soft $\gamma$ radiation 
off light fermions $f=e,\mu,\dots$
and virtual QED corrections. 
In this definition of reduced cross sections [see also 
\cite{Wien}], the logarithmically large QED radiative 
corrections are consistently eliminated in a gauge-invariant way. 
By the same token, the reduced cross sections are defined without
taking into account beamstrahlung.   \\

\item[(c)] \underline{Cross sections for hadron collisions:}\\[1mm]
Cross sections for proton collisions at Tevatron and LHC, $\sigma(pp \to 
\{\tilde{F}\})$, include all QCD and other available 
corrections, with infrared and collinear singularities tamed by
defining inclusive observables, or properly defined jet characteristics, 
and introducing the renormalized parton densities, provided parametrically
by the PDF collaborations \cite{CTEQ,MRST}.   
    
\end{itemize}

\section{TASKS OF THE SPA PROJECT}

A successful reconstruction of the fundamental structure 
of the supersymmetric theory at the high scale 
and the proper understanding of the nature of cold dark matter
from experimental data 
require the precise analysis of all
information that will become available from collider experiments,
low-energy experiments, astrophysical and cosmological observations.  
Preliminary studies [see Sect.5], initiating this SPA Project, 
have shown that
while this aim can in principle be achieved, 
it still needs much additional work both on 
the theoretical as well as on the experimental side.  In particular, 
we identify the following areas of research as central tasks 
of the SPA Project: \\ 

\noindent
\underline{Higher-order calculations}\\[2mm]
\noindent
While the precision of SUSY calculations has gradually shifted 
   from leading-order (LO) to next-to-leading order (NLO) accuracy 
   [and, in some areas, beyond], the
   present level still does not match the expected experimental
   precision, particularly in coherent LHC+ILC analyses.
   The experimental precision, however, has to be fully exploited in 
   order to draw firm conclusions on the fundamental theory. To close 
   this gap, the SPA Project foresees new efforts to push the frontier
   in higher-order SUSY calculations to the line necessary for
   the proper interpretation of experimental analyses. \\

\noindent
\underline{Improving the understanding of the $\drbar$ scheme} \\[2mm]
\noindent
The $\drbar$ scheme recommended for higher-order calculations 
can be formulated in a mathematically consistent way \cite{Stockinger:2005gx}
and is technically
most convenient.  Many explicit checks at the one-loop level have shown that
the $\drbar$ method generates the correct counter terms. However, there is no
complete proof yet that it preserves supersymmetry and gauge invariance in all
cases.
Therefore, as the precision of SUSY calculations
is pushed to higher orders, the SPA Project also requires
further investigation of the symmetry identities in the $\drbar$ scheme. 

Moreover, there is an obvious dichotomy between the $\drbar$ scheme, which is
convenient for the definition of SUSY parameters and their renormalization
group evolution, and the $\msbar$ scheme, which is generally adopted for the
calculation of hadronic processes \cite{Beenakker:1996ch}.
While, as argued before, the $\msbar$ scheme requires {\it{ad-hoc}} counter 
terms to restore supersymmetry, 
in the $\drbar$ scheme a finite shift from the commonly used $\msbar$ 
density functions to the $\drbar$ density functions has to be carried 
out~\cite{37a}.
Moreover, for massive final state particles spurious density functions for the 
$(4 - D)$ gluon components have to be introduced to comply with the 
factorization theorem, see~\cite{37b,37c} for details.
Formulating an efficient combination of the most attractive elements of
both schemes in describing hadronic processes is therefore an important task
of the project.  \\

\noindent
\underline{Improving experimental and theoretical precision} \\[2mm]
\noindent
The set of observables that has been included so far in
   experimental analyses, by no means exhausts the opportunities which
   data at LHC and at ILC are expected to provide in the future. 
   SPA Project studies will aim to
   identify any new channels that can
   give additional information, either independent or 
   redundant [improving fit results],
   and they will include them in a unified framework. In
   connection with realistic estimates of theoretical uncertainties, a
   solid account of error sources and correlations has to be
   achieved. Furthermore, the sophistication of the experimental
   results will be refined by including more precise signal and background 
   calculations, and improved simulations as mandatory for
   the analysis of real data.  \\

\noindent
\underline{Coherent LHC + ILC analyses} \\[2mm]
\noindent
We put particular emphasis on the coherent combination of future
   data obtained at LHC and ILC. While the LHC will most likely discover
   SUSY particles, if they exist, and will allow for the first
   tests of the SUSY paradigm, $e^+e^-$ data make possible high-precision
   investigations of the weakly-interacting sector. Feedback and coherently
   combined analyses, which will greatly benefit from a concurrent
   running of both colliders, are indispensable for a meaningful
   answer to the questions raised in the present context. 
   Studies as initiated by the LHC/LC Study Group \cite{R2B} 
   are a vital part of the SPA Project.  \\

\noindent
\underline{Determining SUSY Lagrangian parameters} \\[2mm]
\noindent
While at leading order the Lagrangian parameters connected
   with different supersymmetric particle sectors can in general be isolated
   and extracted analytically from closely associated observables,
   the analysis is much more complex at higher orders. Higher orders
   introduce the interdependence of all sectors in the observables.
   The development of consistent analyses for the global determination of
   the Lagrangian parameters in this complex situation has started and, 
   conform with general expectations
   for iterative steps in perturbative expansions, they can be carried out
   consistently with as few assumptions as possible. The set of Lagrangian
   parameters and their experimental error matrix can be determined, including 
   higher-order corrections. However, the experimental procedure must still
   be supplemented by corresponding theoretical errors
   and their correlations. \\      

\noindent
\underline{Cold dark matter} \\[2mm]
\noindent
As the precision is refined, astrophysical data play 
   an increasingly important role in confronting supersymmetry with 
   experiments. The class of models
   conserving R-parity predict a weakly interacting,
   massive, stable particle. The relic abundance of this particle 
   imposes crucial limits on supersymmetric scenarios~\cite{Djouadi:2001yk}. 
   While among the
   supersymmetry breaking models versions of mSUGRA and of gaugino 
   mediation \cite{CotBal} have been analyzed in detail, the analyses 
   have to be extended systematically to other scenarios. In  
   models that account for the relic density, specific requirements on 
   the accuracies must be achieved when the CDM particle is studied in 
   high-energy physics laboratory experiments \cite{Puk}. 
   In turn, predictions 
   based on the comprehensive parameter analysis of high-energy 
   experiments determine the cross 
   sections for astrophysical scattering 
   experiments by which the nature of the cold dark matter particles 
   can be established.
   The SPA Project provides a platform for a systematic and continuous 
   interplay between the astrophysics and high-energy physics disciplines
   and the mutual refinement of their programs in the future. \\

\noindent
\underline{Extended SUSY scenarios} \\[2mm]
\noindent
The MSSM, in particular the parameter set SPS1a$'$ that we 
   suggest for a first study, 
   provides a benchmark scenario for
   developing and testing the tools needed for a successful analysis
   of future SUSY data.  However, neither this specific point nor
   the MSSM itself may be the correct model for low-scale SUSY.
   Various parameter sets [for instance other representative
   mSUGRA points as well as non-universal SUGRA, GMSB, AMSB, and other scenarios,
   see Ref.~\cite{HaberPDG} for a brief summary] 
   and extended models have therefore to be investigated within the SPA Project. 
   In particular,  
   models which incorporate the right-handed neutrino sector,
   must be analyzed extensively~\cite{Baer:2000hx}.
   Furthermore,
   $CP$ violation, $R$-parity violation, flavor violation, NMSSM and
   extended gauge groups are among the roads that nature may have
   taken in the SUSY sector. The SPA conventions are formulated
   so generally that they can be applied to all these
   scenarios. The goal of deriving the
   fundamental structure from data will also to be pursued 
   for many facets in this more general context.   

\renewcommand{\arraystretch}{1.2}
\begin{table}[t] \small
\begin{center}
\begin{tabular}{|c|c||c|c|}
  \hline
  Parameter & SM input &
  Parameter & SM input \\ \hline \hline
  $m_e$     & $5.110\cdot 10^{-4}$ & $m_t^{pole}$ & $172.7$  \\
  $m_\mu$   & 0.1057               & $m_b(m_b)$   & $4.2$ \\    
  \cline{3-4} \cline{3-4}
  $m_\tau$  & 1.777                & $m_Z$        & $91.1876$ \\             
  \cline{1-2} 
  $m_u(Q)$  & $3 \cdot 10^{-3}$    & $G_F$        & $1.1664\cdot 10^{-5}$ \\
  $m_d(Q)$  & $7 \cdot 10^{-3}$    & $ 1/\alpha$  & $137.036$ \\ 
  $m_s(Q)$  & 0.12 & $\Delta \alpha^{(5)}_{had}$  & $0.02769$ \\
  $m_c(m_c)$& 1.2  & $\alpha_s^\msbar(m_Z)$       & $0.119$ \\ \hline
\end{tabular}
\end{center}
\caption{\small \it
  Numerical values of the SM input to SPS1a$'$. Masses are given 
  in \GeV, for the leptons and the $t$ quark the pole masses, 
  for the lighter quarks the $\msbar$ masses either at 
  the mass scale itself, for $c$, $b$, or, for $u$, $d$, $s$, 
  at the scale $Q=2$ \GeV.}
\label{tab:sminput}
\end{table}

\section{EXAMPLE: REF POINT SPS1a$'$ }

To test the internal consistency of the SPA 
scheme and to explore the potential of such extended
experimental and theoretical analyses we have defined, as an
example, the CP and R-parity invariant MSSM reference point SPS1a$'$. 
Of course, the
SPA Convention is set up to cover also more general scenarios. 

The results for SPS1a$'$ presented below are based
on preliminary experimental simulations. In some cases,
however, extrapolations from earlier analyses for SPS1a
and other reference points have been used in order to
substitute missing information necessary for a first comprehensive
test of all aspects of the SPA Project. It is
obvious that many detailed simulations are needed to
demonstrate the full power of predicting the fundamental
supersymmetric parameters from future sets of
LHC and ILC data.

In $e^+e^-$ annihilation experimental progress is expected
for the heavy chargino and neutralinos. 
Combining the results of such studies with LHC data appear very promising
and lead to improved mass determinations~\cite{desch}.
New techniques to determine slepton masses from cascade
decays as very narrow resonances \cite{aguilar,berggren} should
be applied. For cross section measurements and other
sparticle properties methods to determine the decay branching ratios should
be developed. At the LHC a recently proposed {mass
relation method} offers substantial improvements in the reconstruction
of squark and gluino masses \cite{kawagoe}. \\

\noindent
\underline{Analysis of SUSY Lagrangian parameters} \\[2mm]
\noindent
The roots defining the Reference Point SPS1a$'$ are the mSUGRA
parameters [in the conventional notation for CMSSM --
see \cite{cmssm} for the tighter original definition]
in the set 
\begin{eqnarray*}
  \fbox{$%
    \begin{array}{lcrlcr}
      M_{1/2} &=& 250 \;\mbox{GeV}  \;\;  &  \mbox{sign}(\mu)      &=& +1
      \\  
      M_0     &=&  70 \;\mbox{GeV}  \;\;  &  \tan\beta (\tilde M)  &=& 10
      \\ 
      A_0     &=&-300 \;\mbox{GeV}  \;\;  &                        & &         
    \end{array}  $} 
  \label{param}
\end{eqnarray*}
The left column, listing the universal gaugino mass $M_{1/2}$,
the scalar mass $M_0$ and the trilinear coupling $A_0$ [Yukawa couplings
factored out], is defined at the GUT scale $M_{\rm GUT}$. The point is
close to the original Snowmass point SPS1a~\cite{Allanach:2002nj}; 
the scalar mass parameter $M_0$ is lowered slightly at the GUT
scale from $100~\GeV$ to $70~\GeV$ and $A_0$ is changed from $-100~\GeV$ to 
$-300~\GeV$.
The values of the SM input parameters are collected in
Table~\ref{tab:sminput}.  
Extrapolation of the above  mSUGRA parameters 
down to the $\tilde{M} = $ 1 TeV scale 
generates the MSSM Lagrangian parameters.  
Table~\ref{tab:par} displays  
the couplings and mass parameters after 
being evolved from $M_{\rm GUT}$ to $\tilde{M}$ using 
the RGE part of the program
{\ttfamily SPheno} \cite{Porod:2003um}
which is based on two-loop analyses of the $\beta$-functions
as well as the other evolution coefficients 
(other codes can be used equally well). 

This SPS1a$'$ set is compatible with all high-energy mass bounds
and with the low-energy precision data, as well as with the observed
CDM data, calculated as
${\cal B}(b \to s\gamma) = 3.0 \cdot 10^{-4}$~\cite{Belanger:2004yn},  
$\Delta [g-2]_\mu/2 = 34 \cdot 10^{-10}$~\cite{Heinemeyer:1998yj},
$\Delta\rho_{\rm SUSY} = 2.1 \cdot 10^{-4}$~\cite{Heinemeyer:1998yj},
and $\Omega_{\rm CDM} h^2 = 0.10$~\cite{Belanger:2004yn}.

\renewcommand{\arraystretch}{1.2}
\begin{table}[t] \small
\begin{center}
$ \begin{array}{|c|c||c|c|}
\hline
  \ \mbox{Parameter}\ 
                      & \ \mbox{SPS1a$'$ value} \ 
& \ \mbox{Parameter}\ 
                      & \ \mbox{SPS1a$'$ value} \  \\ \hline \hline
 g'   &   0.3636  &  M_1       & 103.3  \\
 g    &   0.6479  &  M_2       & 193.2  \\
 g_s  &   1.0844  &  M_3       & 571.7  \\ \hline
 Y_\tau & 0.1034  & A_\tau     & -445.2 \\
 Y_t  &   0.8678  & A_t        & -565.1 \\
 Y_b  &   0.1354  & A_b        & -943.4 \\ \hline
 \mu  &    396.0  & \tan\beta  & 10.0   \\ \hline
 M_{H_d}  & 159.8  & 
 |M_{H_u}| & 378.3   \\ \hline  
 M_{L_1}  & 181.0  &  M_{L_3}  & 179.3  \\
 M_{E_1}  & 115.7  &  M_{E_3}  & 110.0  \\
 M_{Q_1}  & 525.8  &  M_{Q_3}  & 471.4  \\
 M_{U_1}  & 507.2  &  M_{U_3}  & 387.5  \\
 M_{D_1}  & 505.0  &  M_{D_3}  & 500.9  \\ \hline
\end{array} $\\
\end{center} 
\caption{\small \it The $\drbar$ SUSY Lagrangian parameters 
  at the scale $\tilde M=1$~{\TeV} 
  in SPS1a$'$ from \cite{Porod:2003um} [mass unit in \GeV; $M_{H_u}^2$ negative]. 
  In addition,  
  gauge and Yukawa couplings at this scale are given in the 
  $\drbar$ scheme.}
\label{tab:par}
\end{table}

\renewcommand{\arraystretch}{1.2}
\begin{table}[t] \footnotesize
\newcommand{\emptyccc}{\multicolumn{3}{c|}{}}
\begin{center} $
  \begin{array}{|c||c|c|}
    \hline
    \ \mbox{Particle}\ & \ \mbox{Mass}\ [\GeV]\ 
                       & \ \delta_{\rm scale}\ [\GeV]\ \\
    \hline \hline
    h^0        & 116.0 & 1.3 \\
    H^0        & 425.0 & 0.7 \\ \hline
    \nt_1      &  97.7 & 0.4 \\
    \nt_2      & 183.9 & 1.2 \\ 
    \nt_4      & 413.9 & 1.2 \\
    \cx^\pm_1  & 183.7 & 1.3 \\ \hline
    \ser       & 125.3 & 1.2 \\
    \sel       & 189.9 & 0.4 \\
    \stau_1    & 107.9 & 0.5 \\ \hline
    \tilde q_R & 547.2 & 9.4  \\
    \tilde q_L & 564.7 & 10.2 \\
    \st_1      & 366.5 & 5.4  \\
    \sbot_1    & 506.3 & 8.0  \\  \hline
    \sg        & 607.1 & 1.4 \\
    \hline
  \end{array}$
\end{center}
\caption{\small \it Supersymmetric masses for the SUSY scale 
  $\tilde{M} = 1\;{\TeV}$, 
  and their variation if $\tilde{M}$ is shifted 
  to $0.1\;{\TeV}$.} 
\label{tab:scale}
\end{table}

The physical [pole] masses of the supersymmetric particles are presented
in Table~\ref{fig:spec}.
The connection between the Lagrangian parameters and the physical pole
masses is presently encoded at the one-loop level for the masses of
the SUSY particles, and at the two-loop level for the Higgs masses.
QCD effects on the heavy quark masses are accounted for to two-loop
accuracy. 

A systematic comparison with the
other public programs {\ttfamily ISAJET} \cite{Paige:2003mg}, 
{\ttfamily SOFTSUSY} \cite{Allanach:2001kg} 
and {\ttfamily SuSpect} \cite{Djouadi:2002ze}
has been performed in \cite{Allanach:2003jw} to estimate the
technical accuracy that can pres\-ently be reached
in the evolution.
The codes include full two-loop RGEs for all parameters as well as one-loop
formulas for threshold corrections. 
The agreement between the actual versions of
these calculations is in general within one percent. 
A special case are the on-shell masses of the Higgs bosons 
which have been calculated by {\ttfamily FeynHiggs} \cite{Heinemeyer:1998yj} 
starting from the {\ttfamily SPheno} Lagrangian parameters
as input. 
Here, discrepancies for the mass of the lightest Higgs boson
amount to 2\% or more which can 
be attributed to different renormalization schemes (see also 
\cite{Allanach:2004rh} for detailed discussions). 

Besides the comparison between different codes for spectrum calculations, 
a crude internal estimate of the theoretical errors at the
present level of the loop calculations may be obtained by shifting the
matching point $\tilde{M}$ from 1~TeV down to 0.1~TeV. A sample of 
particle mass
shifts associated with such a variation of the SUSY scale parameter 
is displayed in Table~\ref{tab:scale}. With errors at the
percent level, the experimental precision at LHC can be matched in
general. However, it is obvious that another order of magnitude, the
per-mil level, is required in the theoretical precision to match the
expected experimental precision at ILC and in coherent LHC/ILC analyses --
{i.e.}, calculations of the next loop are called for\footnote{With
    $\beta$ functions and evolution coefficients in the RGEs already    
    available to third order \cite{JJK}, the calculation of the two-loop
    order for the relation between the Lagrangian parameters and
    the physical pole masses have been carried out in the approximation of
    massless vector bosons \cite{martin} }.

To perform experimental simulations, the branching ratios of the
decay modes are crucial: these have been calculated using 
{\ttfamily FeynHiggs}~\cite{Heinemeyer:1998yj} 
and {\ttfamily SDECAY}~\cite{sdecay}; 
similar results may be obtained using {\ttfamily CPSuperH}~\cite{Lee:2003nt}.
The most important decay channels of the supersymmetric particles and Higgs
bosons in SPS1a$'$  are collected in the Appendix, while the complete set is
available from the SPA web-site.  
Cross sections for the production of squarks, gluinos, gauginos and sleptons 
at the LHC are presented as a function of mass including the point SPS1a$'$. 
Typical cross sections for pair
production of charginos, neutralinos and sleptons at the ILC are
presented for the point SPS1a$'$ as a function of the collider energy. 

\begin{table*}[t] \small
\setlength{\unitlength}{1mm}
\begin{picture}(170,80)(0,0)
\put(85,5){\epsfig{figure=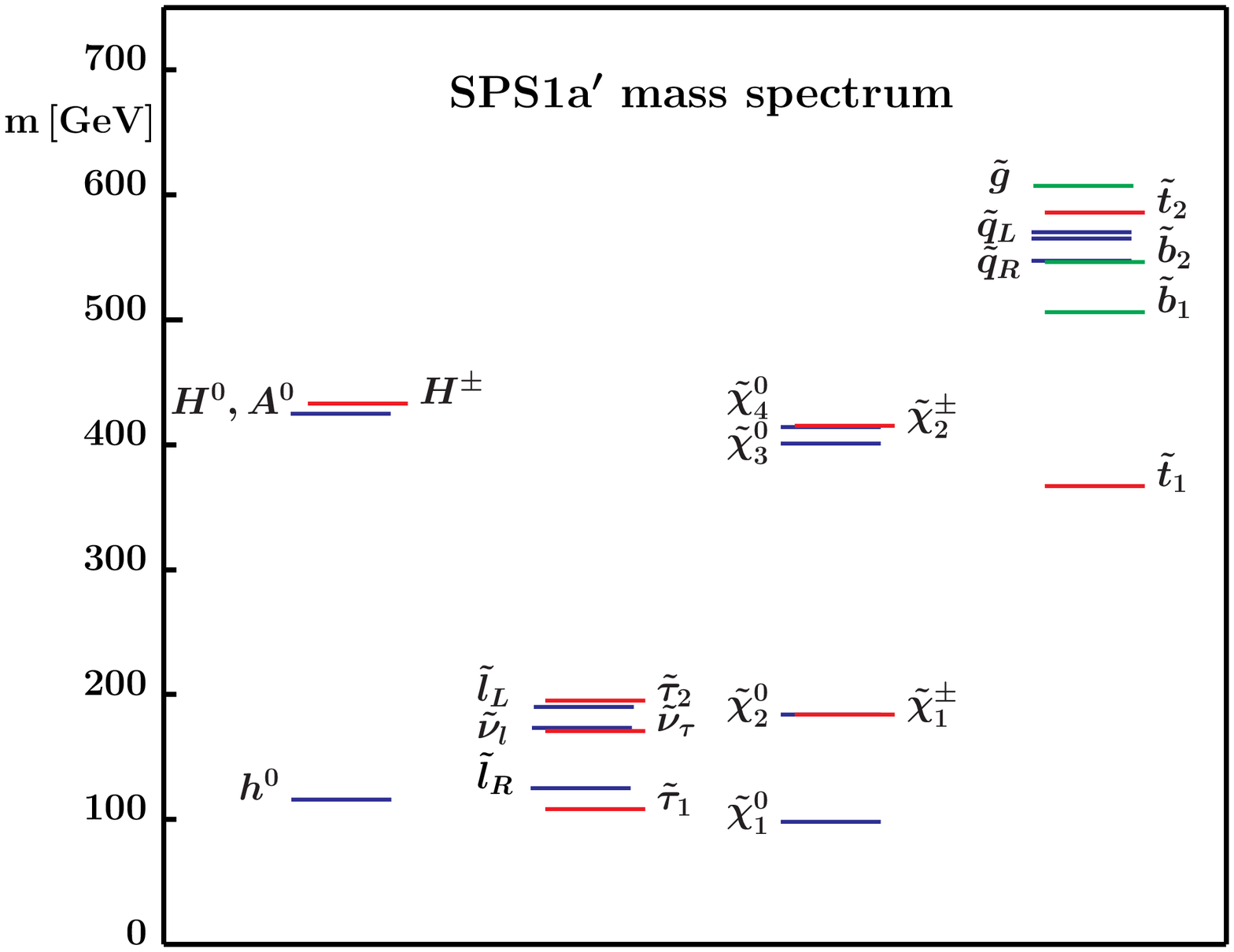,height=75mm,width=80mm}}
\put( 5,42){{ \small $
\renewcommand{\arraystretch}{1.2}
\begin{array}{|c|c||c|c|}
  \hline
  \ \mbox{Particle}\ & \ \mbox{Mass [GeV]}\ & 
  \ \mbox{Particle}\ & \ \mbox{Mass [GeV]}\ \\
  \hline \hline
  h^0    &  116.0 & \stau_1 &  107.9 \\
  H^0    &  425.0 & \stau_2 &  194.9 \\
  A^0    &  424.9 & \snt    &  170.5 \\ \cline{3-4}
  H^+    &  432.7 & \ur     &  547.2 \\ \cline{1-2}
  \nt_1  &   97.7 & \ul     &  564.7 \\
  \nt_2  &  183.9 & \dr     &  546.9 \\
  \nt_3  &  400.5 & \dl     &  570.1 \\ \cline{3-4}
  \nt_4  &  413.9 & \st_1   &  366.5 \\
  \cp_1  &  183.7 & \st_2   &  585.5 \\
  \cp_2  &  415.4 & \sbot_1 &  506.3 \\ \cline{1-2}\cline{1-2}
  \ser   &  125.3 & \sbot_2 &  545.7 \\ \cline{3-4}
  \sel   &  189.9 & \sg     &  607.1 \\
  \sne   &  172.5 &         &        \\
  \hline
\end{array}    $ }}
\end{picture}
\caption{\small \it Mass spectrum of
  supersymmetric particles \cite{Porod:2003um} and
  Higgs bosons \cite{Heinemeyer:1998yj} in the reference point SPS1a$'$.
  The masses in the second generation coincide
  with the first generation.}
\label{fig:spec}
\end{table*}

\renewcommand{\arraystretch}{1.2}
\begin{table}[h] \footnotesize
\begin{center} $
  \begin{array}{|c|c||c|c||c|}
    \hline 
    \ \mbox{Particle} \ &
    \ \ \mbox{Mass}\ \  & \ \mbox{``LHC''}\ & \ \mbox{``ILC''}\ 
                        & \ \mbox{``LHC+ILC''}\ \\ 
    \hline\hline
    h^0                 & 116.0 & 0.25 & 0.05 & 0.05 \\
    H^0                 & 425.0 &      & 1.5  & 1.5  \\
    \hline 
    \tilde{\chi}^0_1    &  97.7 & 4.8  & 0.05 & 0.05 \\
    \tilde{\chi}^0_2    & 183.9 & 4.7  & 1.2  & 0.08 \\
    \tilde{\chi}^0_4    & 413.9 & 5.1  & 3-5  & 2.5  \\
    \tilde{\chi}^\pm_1  & 183.7 &      & 0.55 & 0.55 \\ \hline 
    \tilde{e}_R         & 125.3 & 4.8  & 0.05 & 0.05 \\
    \tilde{e}_L         & 189.9 & 5.0  & 0.18 & 0.18 \\
    \tilde{\tau}_1      & 107.9 & 5-8  & 0.24 & 0.24 \\ \hline
    \tilde{q}_R         & 547.2 & 7-12 & -    & 5-11 \\
    \tilde{q}_L         & 564.7 & 8.7  & -    & 4.9  \\
    \tilde{t}_1         & 366.5 &      & 1.9  & 1.9  \\
    \tilde{b}_1         & 506.3 & 7.5  & -    & 5.7  \\ \hline
    \tilde{g}           & 607.1 & 8.0  & -    & 6.5  \\ \hline
  \end{array}$ \\
\end{center}
\caption{ \small \it Accuracies for representative mass measurements
    of SUSY particles in individual LHC, ILC and 
    coherent ``LHC+ILC'' analyses
    for the reference point SPS1a$'$ [mass units in {\GeV}].
    $\tilde q_R$ and $\tilde q_L$ represent the flavors
    $q=u,d,c,s$. [Errors presently extrapolated from SPS1a simulations.] }
\label{tab:massesA}
\end{table}

If SPS1a$'$, or a SUSY parameter set in the range of similar
mass scales, is realized in nature, a plethora of
interesting channels can be exploited to extract the basic
supersymmetry parameters when combining experimental information from
sharp edges in mass distributions at LHC with measurements of decay
spectra and threshold excitation curves at an $e^+e^-$ collider with
energy up to 1 TeV {\cite{Allanach:2004ed}}. 
From the simulated experimental errors the data
analysis performed coherently for the two machines gives rise to a
very precise picture of the supersymmetric particle spectrum as
demonstrated in Table~\ref{tab:massesA}.

While the picture so far had been based on evaluating the experimental
observables channel by channel, global analysis programs 
have become available \cite{Lafaye:2004cn,fittino} in 
which the whole set of data, masses, cross sections, branching ratios,
{etc.} is exploited coherently to extract the
Lagrangian parameters in the optimal way after including the available
radiative corrections for masses and cross sections. 
With increasing numbers of observables the analyses can be expanded
and refined in a systematic way.
The present quality of such an analysis~\cite{fittino} 
can be judged from the results shown in Table~\ref{tab:params}. 
These errors are purely experimental and do
not include the theoretical counterpart which must be improved
considerably before matching the experimental standards. \\

\begin{figure*}[t]
\begin{center}
\setlength{\unitlength}{1mm}
\begin{picture}(150,75)(5,10)
  \put( 0,-70){\epsfig{figure=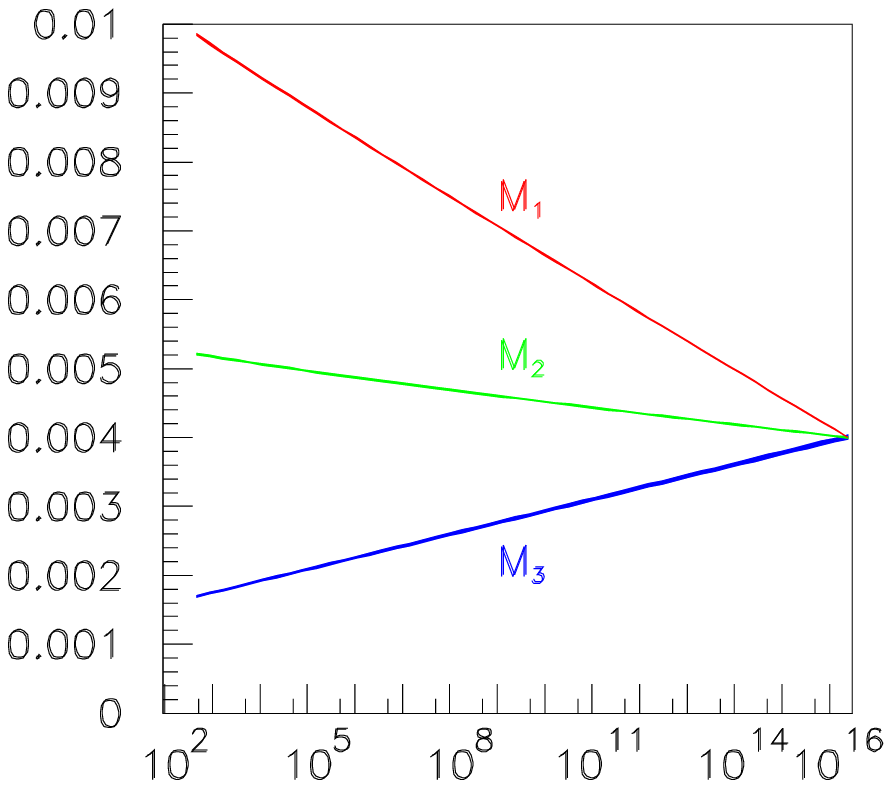,width=165mm}}
  \put(85,-70){\epsfig{figure=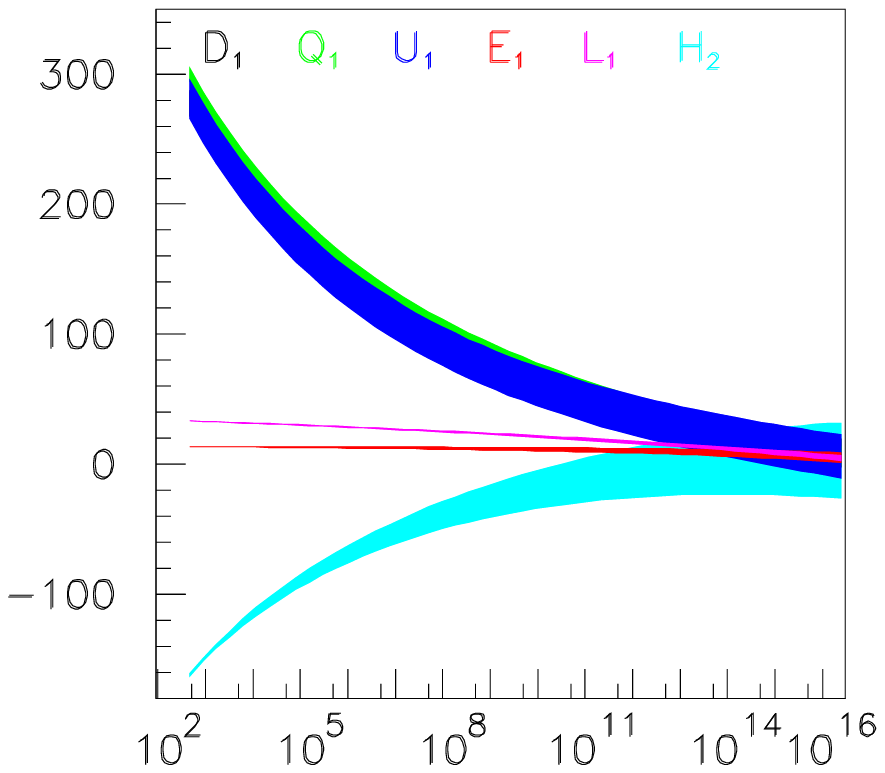,width=165mm}}
  \put(44, 30){\epsfig{figure=box.eps,height=7mm,width=8mm}}
  \put(44, 48){\epsfig{figure=box.eps,height=7mm,width=8mm}}
  \put(44, 62){\epsfig{figure=box.eps,height=7mm,width=8mm}}
  \put(-5,53){\begin{sideways}\large $1/M_i~[\GeV^{-1}]$ \end{sideways}}
  \put(85,52){\begin{sideways}\large $M^2_{\tilde f}~[10^3\,\GeV^2]$ \end{sideways}}
  \put( 60,10){\large $Q~[\GeV]$}
  \put(145,10){\large $Q~[\GeV]$}
  \put(44, 32){\color{blue}\large $M_3^{-1}$ }
  \put(44, 49){\color{green}\large $M_2^{-1}$ }
  \put(44, 63){\color{red}\large $M_1^{-1}$ }
\end{picture}
\end{center}
\caption{\small \it 
  Running of the gaugino and scalar mass parameters as a function of the 
  scale $Q$ in SPS1a$'$  \cite{Porod:2003um}.
  Only experimental errors are taken into account; theoretical errors are
  assumed to be reduced to the same size in the future.}
\label{fig:running}
\end{figure*}

\noindent
\underline{Extrapolation to the GUT scale} \\[2mm]
\noindent
Based on the parameters extracted at the scale $\tilde{M}$, we can
approach the reconstruction of the fundamental supersymmetric theory
and the related microscopic picture of the mechanism breaking 
supersymmetry. The experimental information is exploited to the maximum
extent possible in the bottom-up approach \cite{Blair:2000gy} 
in which the extrapolation from
$\tilde{M}$ to the GUT/Planck scale is performed by the renormalization group
evolution for all parameters,
with the GUT scale defined by the unification point of the two electroweak couplings.
In this approach the calculation of loops
and $\beta$ functions governing the extrapolation to the high scale is
based on nothing but experimentally measured parameters. Typical examples
for the evolution of the gaugino and scalar mass parameters are
presented in Fig.~\ref{fig:running}. While the determination of the
high-scale parameters in the gaugino/higgsino sector, as well as in
the non-colored slepton sector, is very precise, 
the picture of the colored scalar and Higgs sectors
is still coarse, and strong efforts should be made to refine it considerably. 

\renewcommand{\arraystretch}{1.2}
\begin{table}[t] \footnotesize
\begin{center}
  \begin{tabular}{|c||c|c|}
    \hline
    Parameter & SPS1a$'$ value & Fit error [exp] \\
    \hline\hline
    $M_1$                  & 103.3  &   $ 0.1  $  \\
    $M_2$                  & 193.2  &   $ 0.1  $  \\
    $M_3$                  & 571.7  &   $ 7.8  $  \\
    $\mu$                  & 396.0  &   $ 1.1  $  \\ \hline
    $M_{L_1}$              & 181.0  &   $ 0.2   $  \\
    $M_{E_1}$              & 115.7  &   $ 0.4   $  \\
    $M_{L_3}$              & 179.3  &   $ 1.2   $   \\ \hline
    $M_{Q_1}$              & 525.8  &   $ 5.2   $   \\
    $M_{D_1}$              & 505.0  &   $ 17.3 \;\; $   \\
    $M_{Q_3}$               & 471.4  &   $ 4.9   $    \\ \hline
    $m_{\mathrm{A}}$       & 372.0  &   $ 0.8  $   \\
    $A_{\mathrm{t}}$       &--565.1~&   $ 24.6 \;\,  $   \\
    $\tan\beta$            & ~10.0  &   $ 0.3 $  \\ \hline
  \end{tabular}
\end{center}
\caption{\small \it 
  Excerpt of extracted SUSY Lagrangian mass and Higgs parameters 
  at the supersymmetry scale $\tilde{M} = 1\;\TeV$ in the reference point 
  SPS1a$'$ [mass units in {\GeV}]. } 
\label{tab:params}
\end{table}

On the other hand, if the 
structure of the theory at the high scale was known {\it a priori} and
merely the experimental determination of the high-scale parameters were 
lacking, then the 
top-down approach would lead to a very precise parametric picture at the
high scale. This is apparent from the fit of the mSUGRA parameters in
SPS1a$'$ displayed in Table~\ref{tab:univ_params}~\cite{Lafaye:2004cn}. 
A high-quality fit of the parameters is a necessary condition, of course, 
for the theory to be correct -- however it is not a sufficient condition; 
deviations from the
theory may hide in large errors of some observables which do not spoil
the quality of the fit in the top-down approach but which are manifest 
in the bottom-up approach. 

\renewcommand{\arraystretch}{1.2}
\begin{table}[htb]
\begin{center} \footnotesize
  \begin{tabular}{|c||c|c|}
    \hline
    Parameter       &  SPS1a$'$ value           & Experimental error \\ 
    \hline\hline
    $M_{\rm GUT}$   &  $2.47 \cdot 10^{16}$ GeV & $0.02 \cdot 10^{16}$ GeV  \\
    $\alpha_{\rm GUT}^{-1}$ &  24.17                    &  0.06      \\ \hline
    \hline
    $M_\frac{1}{2}$ & {\phantom{-}}250 GeV      &  0.2  GeV    \\
    $M_0$           & {\phantom{-0}}70 GeV      &  0.2  GeV     \\
    $A_0$           & -300 GeV                  &  13.0 GeV $\;$  \\   \hline 
    \hline
    $\mu$           &  396.0 GeV                &  0.3 GeV    \\
    $\tan\beta $    &  10                       &  0.3    \\  \hline
  \end{tabular}
\end{center}
\caption{\small \it Comparison of the ideal parameters with the
  experimental expectations in the top-down approach~\cite{fittino}.} 
\label{tab:univ_params}
\end{table}

\vspace{\baselineskip}
\noindent
\underline{Cold dark matter} \\[2mm]
\noindent
Constraints on SUSY cold dark matter can be obtained at LHC 
by specifying the underlying scenario and analyzing all data simultaneously
within the given benchmark model.
From a study of the SPS1a point, based on very large statistics \cite{tovey}, 
one may expect that the
relic density can be determined to $\sim 6$\% for the SPS1a$'$ scenario. 
For SPS1a$'$, the relic density depends on the parameters of the
neutralino and sfermion sector as the dominant channels are
annihilation of neutralinos into fermion pairs and coannihilation with
staus. In particular, for the most sensitive component, coannihilation processes,
the relic density is essentially given by the mass
difference between the lightest slepton $\tilde{\tau}_1$ and the LSP
$\tilde{\chi}^0_1$, which can be directly measured at the ILC. Studies of    
$\tilde{\tau}_1$ production at threshold \cite{bambade} and decay spectra 
to $\nt_1$ in
the continuum \cite{martyn} suggest that for SPS1a$'$, even with moderate
luminosity, a precision of $\sim 2$\% on the cold dark matter abundance is
achievable.  
A systematic analysis of various scenarios is being carried out in the 
LCC project~\cite{lcc} as well as by other groups.

\section{SUMMARY AND OUTLOOK}

If supersymmetry is realized in Nature, future experiments
at the LHC and the ILC will provide very precise measurements
of supersymmetric particle spectra and couplings.
On the theoretical side these measurements must
be matched by equally precise theoretical calculations
and numerical analysis tools. The SPA Project, a joint
theoretical and experimental effort, aims at providing
\begin{itemize}
  \item a well-defined framework for SUSY calculations and
    data analyses,
  \item  all necessary theoretical and computational tools,
  \item  a testground scenario SPS1a$'$,
  \item a platform for future extensions and developments.
\end{itemize}

On this basis coherent analyses of experimental data
can be performed and the fundamental supersymmetric
Lagrangian parameters can be extracted. They can
serve as a firm base for extrapolations to high scales
so that the ultimate supersymmetric theory and the supersymmetry
breaking mechanism can be reconstructed
from future data.

Much work is still needed on the experimental and
theoretical side to achieve these goals at the desired
level of accuracy. Some of the short- and long-term subprojects
have been identified and should be pursued in
the near future.

The SPA Project is a dynamical system expected to
evolve continuously. The current status of the SPA Project,
names of the conveners responsible for specific tasks as
well as links to the available calculational tools, can
be found at the SPA home page \\
 {\ttfamily http://spa.desy.de/spa/}.


\vspace*{27\baselineskip}


\appendix
\section*{APPENDIX}
\subsection*{(a) Decays of Higgs and SUSY particles in SPS1a$'$}
The branching ratios of
Higgs bosons and SUSY particles exceeding 2\% are presented 
in Tables \ref{tab:higgsmodes}--\ref{tab:sq1modes}.
The complete listing including all decays is available on
the SPA web-site  {\ttfamily http://spa.desy.de/spa/}.

\renewcommand{\arraystretch}{1.2}
\begin{table}[h] \centering \footnotesize
  $\begin{array}{|c||c||cc|cc|} 
   \hline & & & & & \\[-2.ex]
   \ \mbox{Higgs}\ & \ m,\Gamma\,[\GeV] \ 
                 & \ \mbox{decay} \ & \ \ \ \mathcal{B} \ \ \
                 & \ \mbox{decay} \ & \ \ \ \mathcal{B} \ \ \
   \\[.5ex] \hline \hline
   h^0      &  116.0 & \tau^-   \tau^+    & 0.077 & W      W^*    & 0.067 \\
            &   4 \times 10^{-3}       
                     & b        \bar b    & 0.773 & g      g      & 0.055\\
            &        & c        \bar c    & 0.021 &               &       \\ \hline
   H^0      &  425.0 & \tau^-   \tau^+    & 0.076 & \nt_1 \nt_2  & 0.038\\
            &  1.2   & b        \bar b    & 0.694 & \nt_2 \nt_2  & 0.020\\
            &        & t        \bar t    & 0.052 & \cp_1 \cm_1  & 0.050\\
            &        & \stau^\pm_1 \stau^\mp_2 & 0.030 &         &      \\ \hline
   A^0      &  424.9 & \tau^-   \tau^+     & 0.057 & \nt_1 \nt_2  & 0.054\\
            & 1.6    & b        \bar b     & 0.521 & \nt_2 \nt_2  & 0.060\\
            &        & t        \bar t     & 0.094 & \cp_1 \cm_1  & 0.163\\
            &        & \stau^\pm_1 \stau^\mp_2 & 0.036 &          &      \\ \hline
   H^+      &  432.7 & \nu_\tau \tau^+     & 0.104 & \cp_1  \nt_1 & 0.143 \\
            &  0.9   & t        \bar b     & 0.672 
                     & \snt \stau^+_1      & 0.071 \\ \hline  
   \end{array} $ \\[1em]
   \caption{\small \it Higgs masses and branching ratios
     $\mathcal{B}>2\%$ in SPS1a$'$ from \cite{Heinemeyer:1998yj}.}
   \label{tab:higgsmodes}
\end{table}

\renewcommand{\arraystretch}{1.2}
\begin{table}[h] \centering \footnotesize
  $\begin{array}{|c||c||cc|cc|}
   \hline & & & & & \\[-2.ex]
   \ \ \tilde{\chi}\ \ & \ m,\Gamma\,[\GeV] \ 
                   & \ \mbox{decay} \ & \ \ \ \mathcal{B} \ \ \
                   & \ \mbox{decay} \ & \ \ \ \mathcal{B} \ \ \                
   \\[.5ex] \hline \hline
   \nt_1    &   97.7 &                      &       &                &       \\ \hline
   \nt_2    &  183.9 & \ser^\pm e^\mp       & 0.025 & \sne \nu_e     & 0.116 \\
            &  0.083 & \stau_1^\pm\tau^\mp  & 0.578 & \snt \nu_\tau  & 0.152 \\ \hline
   \nt_3    &  400.5 & \cx_1^\pm W^\mp      & 0.582 & \nt_1    Z^0   & 0.104 \\
            &    2.4 &                      &       & \nt_2    Z^0   & 0.224
   \\ \hline
   \nt_4    &  413.9 & \stau_2^\pm \tau^\mp & 0.033 & \cx_1^\pm W^\mp& 0.511 \\
            &    2.9 & \sne\nu_e            & 0.042 & \nt_1    Z^0   & 0.022 \\
            &        & \snt\nu_\tau         & 0.042 & \nt_2    Z^0   & 0.024 \\
            &        &                      &       & \nt_1    h^0   & 0.070 \\
            &        &                      &       & \nt_2    h^0   & 0.165 \\ \hline 
   \hline
   \cp_1    &  183.7 & \stau^+_1 \nu_\tau   & 0.536 & \snt \tau^+  & 0.185 \\
            &  0.077 &                      &       & \sne e^+     & 0.133 \\ \hline
   \cp_2    &  415.4 & \sel^+    \nu_e      & 0.041 & \nt_1 W^+    & 0.063 \\
            &    3.1 & \stau^+_2 \nu_\tau   & 0.046 & \nt_2 W^+    & 0.252 \\
            &        & \st_1 b              & 0.109 & \cp_1 Z^0    & 0.221 \\
            &        &                      &       & \cp_1 h^0    & 0.181\\ \hline
   \end{array}$ \\[1em]
   \caption{ \small \it
     Neutralino and chargino masses, widths and branching ratios
     $\mathcal{B}>2\%$ in SPS1a$'$ from \cite{sdecay}; branching ratios for 
     the second generation are the same as for the first generation.}
   \label{tab:chimodes}
 \end{table}

\renewcommand{\arraystretch}{1.2}
\begin{table}[h] \centering \footnotesize
  $\begin{array}{|c||c||cc|cc|}
    \hline & & & & & \\[-2.ex]
    \ \ \tilde{\ell}\ \ & \ m,\Gamma\,[\GeV] \ 
                   & \ \mbox{decay} \ & \ \ \ \mathcal{B} \ \ \
                   & \ \mbox{decay} \ & \ \ \ \mathcal{B} \ \ \              
    \\[.5ex] \hline \hline
    \ser     &  125.3 & \nt_1    e^-     & 1.000 &                &\\
             &   0.10 &                  &       &                &\\ \hline
    \sel     &  189.9 & \nt_1    e^-     & 0.925 & \cm_1 \nu_e    & 0.049 \\
             &   0.12 & \nt_2    e^-     & 0.026 &                & \\  \hline
    \sne     &  172.5 & \nt_1    \nu_e   & 1.000 &                &\\
             &   0.12 &                  &       &                &\\ \hline 
    \stau_1  &  107.9 & \nt_1    \tau^-  & 1.000 &                &\\
             &  0.016 &                  &       &                &\\ \hline
    \stau_2  &  194.9 & \nt_1    \tau^-  & 0.868 & \cm_1 \nu_\tau & 0.086 \\
             &  0.18  & \nt_2    \tau^-  & 0.046 &                &\\ \hline
    \snt     &  170.5 & \nt_1    \nu_\tau& 1.000 &                &\\
             &  0.12  &                  &       &                &\\ \hline
  \end{array}$ \\[1em]
 \caption{\small \it Slepton masses, widths and branching ratios
   $\mathcal{B}>2\%$ in SPS1a$'$ from \cite{sdecay}; 
   branching ratios for 
   the second generation are the same as for the first generation.}
 \label{tab:slmodes}
\end{table}

\renewcommand{\arraystretch}{1.2}
 \begin{table}[h] \centering \footnotesize
  $\begin{array}{|c||c||cc|cc|}
   \hline & & & & & \\[-2.ex]
   \ \ \tilde{q}\ \  & \ m,\Gamma\,[\GeV] \ 
                   & \ \mbox{decay} \ & \ \ \ \mathcal{B} \ \ \
                   & \ \mbox{decay} \ & \ \ \ \mathcal{B} \ \ \            
   \\[.5ex] \hline \hline
   \ur      &  547.2 & \nt_1 u & 0.990 &              &\\
            &  1.2   &         &       &              &\\ \hline 
   \ul      &  564.7 & \nt_2 u & 0.322 & \cp_1 \bar d & 0.656 \\
            &  5.5   &         &       &              &\\ \hline 
   \dr      &  546.9 & \nt_1 d & 0.990 &              &\\
            &  0.3   &         &       &              &\\ \hline
   \dl      &  570.1 & \nt_2 d & 0.316 & \cm_1 \bar u & 0.625 \\
            &  5.4   &         &       &              &       \\ \hline 
   \st_1    &  366.5 & \nt_1 t & 0.219 & \cp_1 b      & 0.719 \\
            &  1.5   & \nt_2 t & 0.062 &              & \\ \hline
   \st_2    &  585.5 & \nt_1 t & 0.042 & \cp_1 b      & 0.265 \\
            &   6.3  & \nt_2 t & 0.103 & \cp_2 b      & 0.168 \\
            &        &         &       & \st_1 Z^0    & 0.354 \\
            &        &         &       & \st_1 h^0    & 0.059 \\ \hline 
   \sbot_1  &  506.3 & \nt_1 b & 0.037 & \cm_1 t     & 0.381 \\
            &   4.4  & \nt_2 b & 0.295 & \st_1 W^-   & 0.281 \\ \hline
   \sbot_2  &  545.7 & \nt_1 b & 0.222 & \cm_1 t     & 0.178 \\
            &   1.0  & \nt_2 b & 0.131 & \st_1 W^-   & 0.401 \\
            &        & \nt_3 b & 0.028 &             &\\
            &        & \nt_4 b & 0.038 &             &\\ \hline
   \hline
   \sg      &  607.1 & \ur   \bar u & 0.086 & \st_1 \bar t   & 0.189 \\ 
            &   5.5  & \ul   \bar u & 0.044 & \sbot_1 \bar b & 0.214 \\ 
            &        & \dr   \bar d & 0.087 & \sbot_2 \bar b & 0.096 \\
            &        & \dl   \bar d & 0.034 &                & \\
   \hline
   \end{array}$ \\[1em]
   \caption{\small \it 
     Masses, widths and branching ratios  $\mathcal{B}>2\%$
     of colored SUSY particles in SPS1a$'$ from \cite{sdecay}; 
     branching ratios for the 
     second generation are the same as for the first generation.} 
   \label{tab:sq1modes}
   \vspace*{-10\baselineskip}
 \end{table}


\vspace*{10\baselineskip}
\subsection*{(b) LHC and ILC cross sections in SPS1a$'$} 

\begin{figure}[h] \vspace*{-3mm}
  \includegraphics[bbllx=150pt,bblly=540pt,bburx=480pt,bbury=780pt,
        width=.5\textwidth]{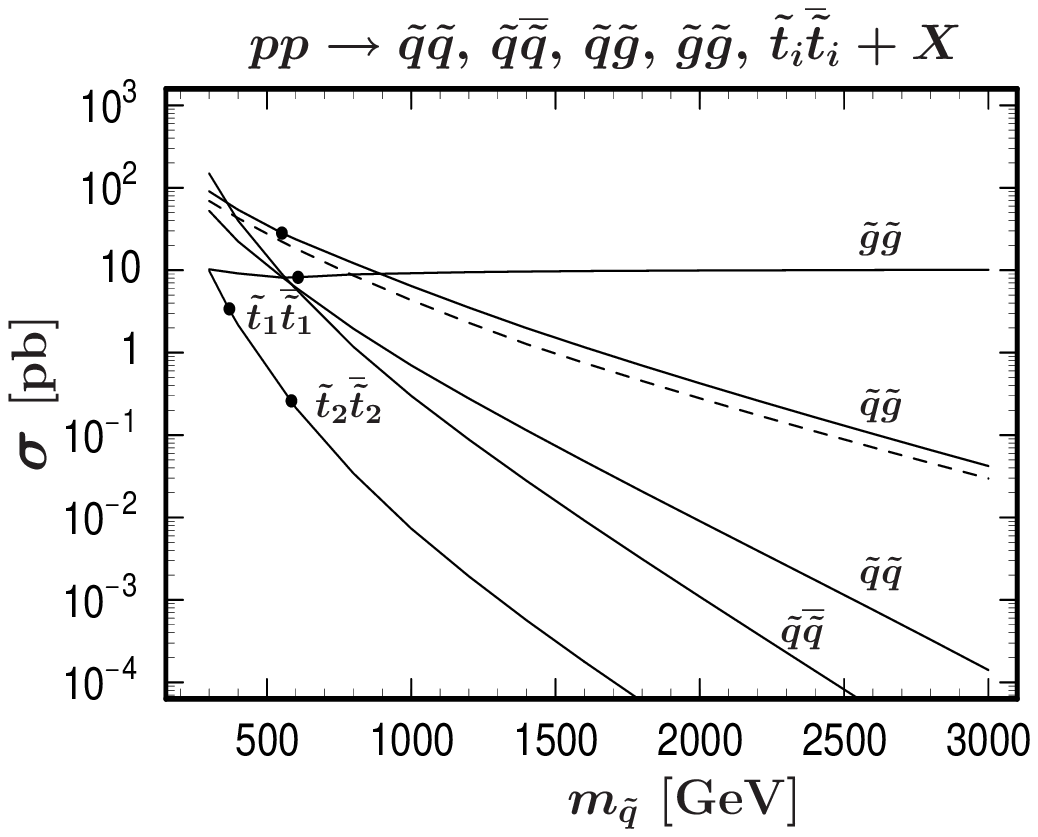}  
  \includegraphics[bbllx=150pt,bblly=540pt,bburx=480pt,bbury=780pt,
        width=.5\textwidth]{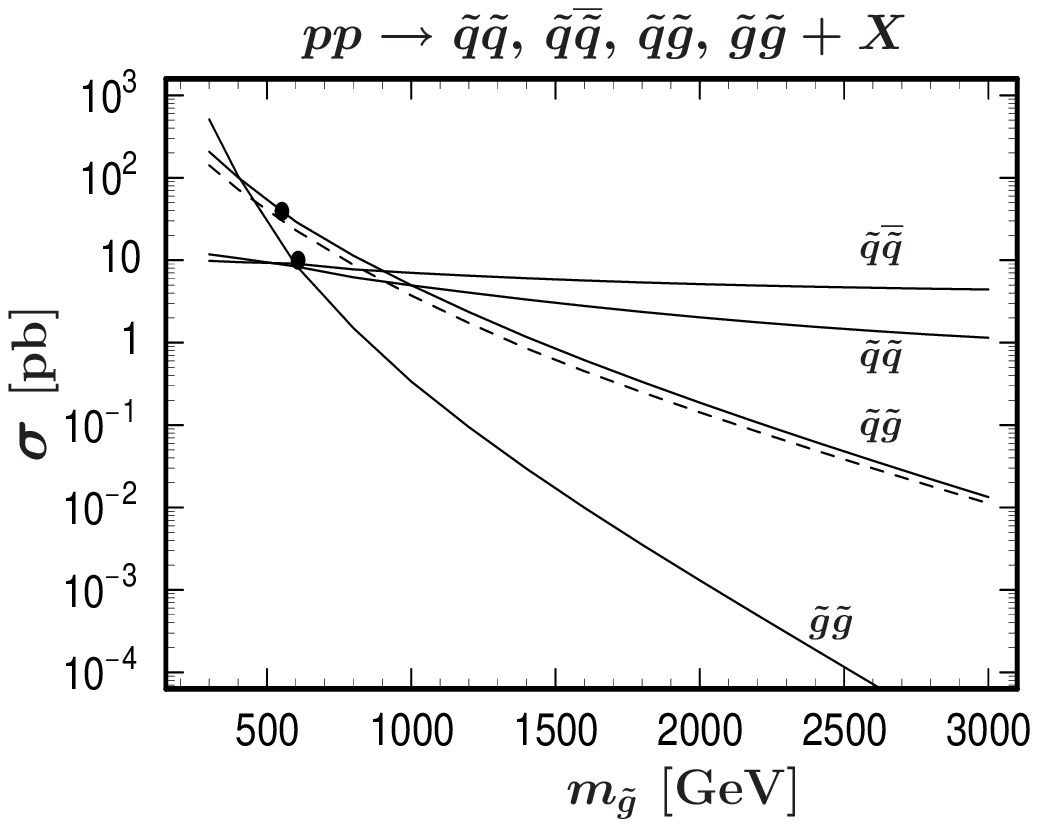} 
  \includegraphics[bbllx=150pt,bblly=540pt,bburx=480pt,bbury=780pt,
        width=.5\textwidth]{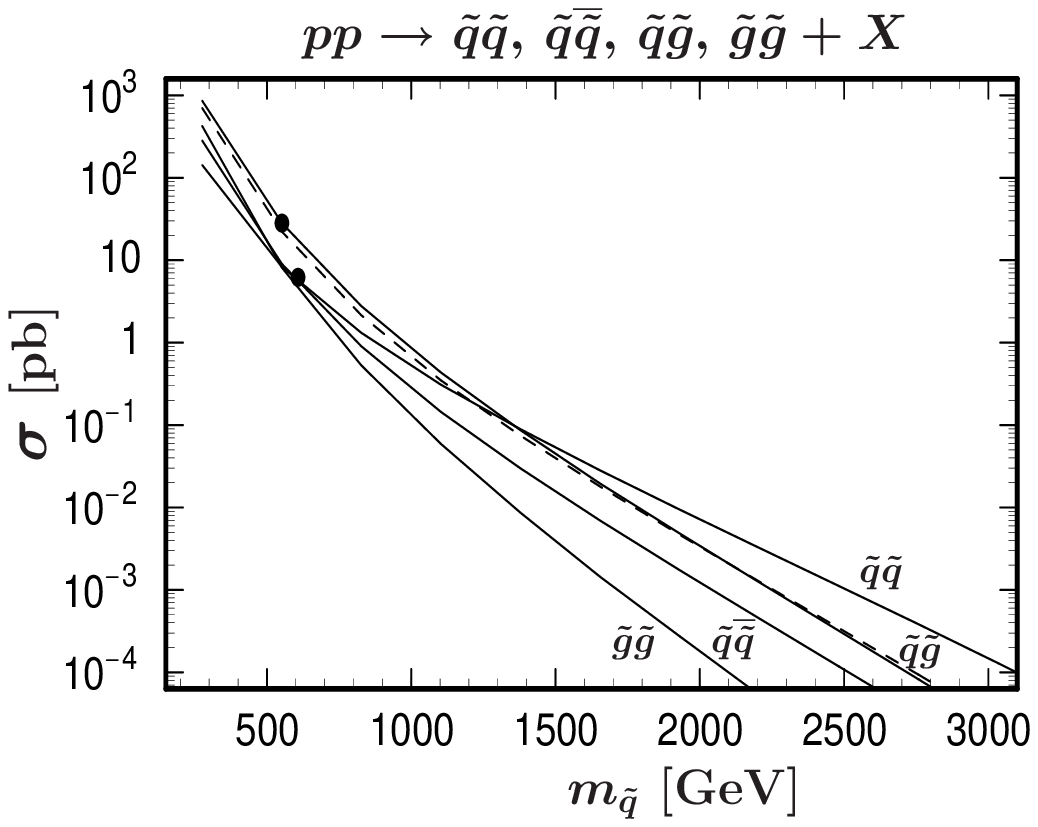}  
  \caption[]{\it Total cross sections for squark and gluino 
    pair production at the LHC~\cite{Beenakker:1996ch,prospino}
    for fixed gluino mass (top), squark mass (center), and
    gluino/squark mass ratio (bottom)
    [fixed parameters corresponding to SPS1a$'$ values].
    Black circles indicate the SPS1a$'$ mass values.
    The Born cross sections (broken lines) 
    are shown for some channels. }
  \label{fig:lhctot1}
\end{figure}

\begin{figure}[p] \centering
  \includegraphics[bbllx=150pt,bblly=540pt,bburx=480pt,bbury=780pt,
        width=.5\textwidth]{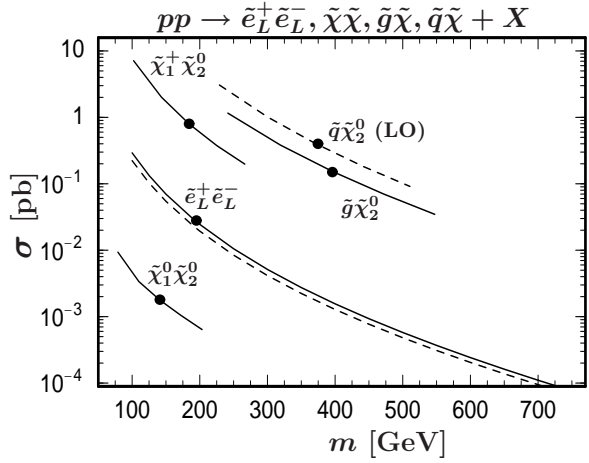}
  \caption[]{\it Generic examples of total cross sections 
    (Drell-Yan and Compton production)
    as a function of the average mass
    for production of sleptons, charginos and neutralinos  
    at the LHC~\cite{Beenakker:1996ch,prospino}.
    The Born cross sections (broken line) are shown for comparison.}  
  \label{fig:lhctot2}
\end{figure}

\begin{figure}[p] 
  \includegraphics[bbllx=150pt,bblly=540pt,bburx=480pt,bbury=780pt,
        width=.5\textwidth]{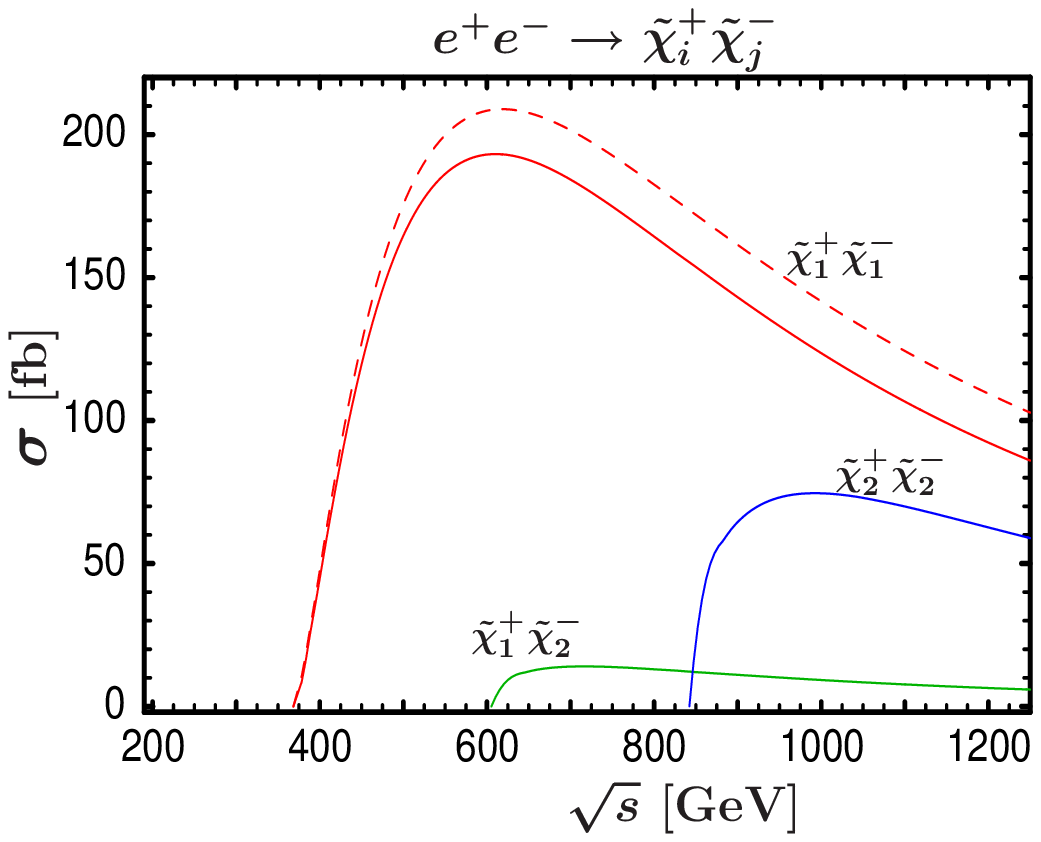} \\[.8em]
  \includegraphics[bbllx=150pt,bblly=540pt,bburx=480pt,bbury=780pt,
        width=.5\textwidth]{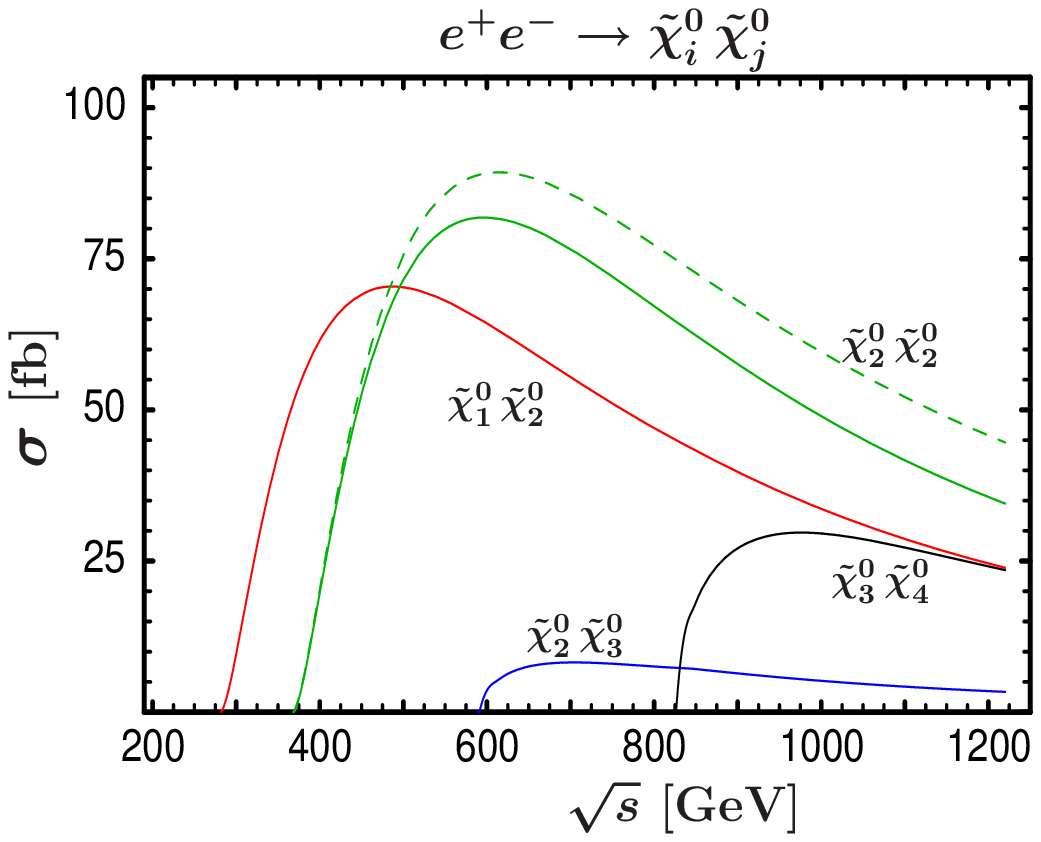}
  \caption{\it Total cross section sections for chargino and neutralino 
      pair production in $e^+e^-$ annihilation \cite{Oller:2005xg}.
      The Born cross sections (broken lines) are shown for a few 
      channels.}   
  \label{fig:ilctot1}
\end{figure}

\begin{figure}[p]
  \includegraphics[bbllx=150pt,bblly=540pt,bburx=480pt,bbury=780pt,
        width=.5\textwidth]{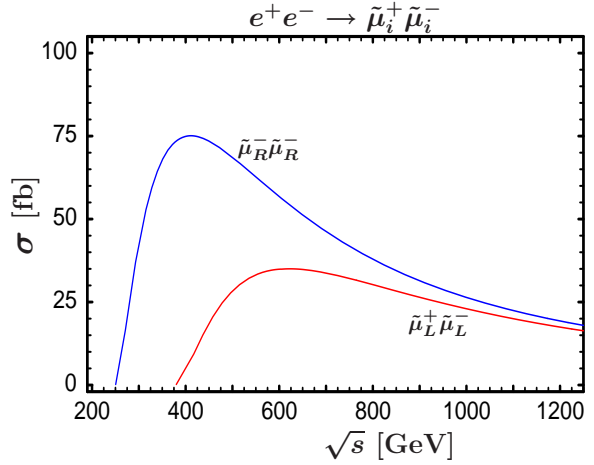} \\[.8em]
  \includegraphics[bbllx=150pt,bblly=540pt,bburx=480pt,bbury=780pt,
        width=.5\textwidth]{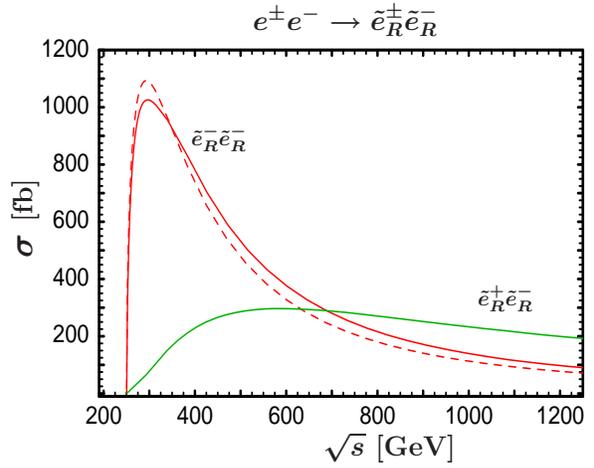}
  \caption{\it Total cross sections for smuon and selectron pair 
      production in $e^{\pm}e^-$ annihilation \cite{Freitas:2003yp}.
      The Born cross section (broken lines) is shown for comparison.}   
  \label{fig:ilctot2}
\end{figure}

\begin{figure}[p]
  \includegraphics[bbllx=150pt,bblly=540pt,bburx=480pt,bbury=780pt,
        width=.5\textwidth]{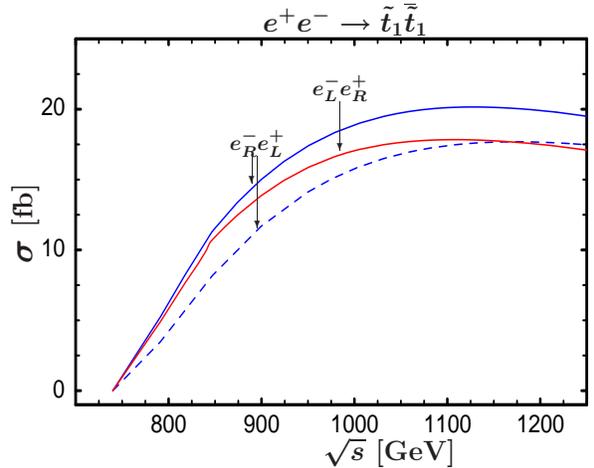}
  \caption{\it Total cross sections for $\tilde{t}_1 \bar{\tilde{t}}_1$ 
      pair production in $e^+e^-$ annihilation
      for left- and right-handed polarized electron ($P_{e^-} = 0.8$) and
      positron ($P_{e^+} = 0.6$) beams \cite{Kovarik:2005ph}.
      The Born cross section (broken line) is shown for comparison.}  
  \label{fig:ilctot3} 
\end{figure}

Total cross sections are presented 
in Figs. \ref{fig:lhctot1} -- \ref{fig:ilctot3}
for SUSY particle production at the LHC 
and the ILC.
%

\end{document}